%% file: bc0343.tex
\documentclass[aps,prl,preprint,tightenlines,superscriptaddress,showpacs,byrevtex,floatfix]{revtex4}

\usepackage{epsfig, amssymb, citesort}
\draft 

\begin{document}

\preprint{BELLE-CONF-0343}

\title{Measurement of the angle {\boldmath$\phi_3$} with
Dalitz analysis of three-body {\boldmath $D^0$} decay from 
{\boldmath $B\to D^0 K$} process}

\input{author.tex}

\date{August 13, 2003}

\begin{abstract} 
A new method to measure the angle $\phi_3$ of the CKM unitarity triangle 
using a Dalitz analysis of the three-body decay of the $D^0$ meson from 
the $B\to D^0 K$ process
has been introduced recently. The method employs the interference 
between $D^0$ and $\bar{D^0}$ to extract both the weak and strong phases. 
We present a study of statistical and systematic errors of this method
using Monte Carlo simulation. A fit of the $D^0$ Dalitz 
distribution has been performed based on a 140 fb$^{-1}$ sample collected by 
the Belle detector. 
We observe the indication of a CP asymmetry with the significance of 2.4 
standard deviations.
The 90\% confidence interval obtained for the $\phi_3$ value is 
$61^{\circ}<\phi_3<142^{\circ}$. 
\end{abstract}


\maketitle

\section{Introduction}

The determination of the Cabibbo-Kobayashi-Maskawa (CKM) matrix elements
\cite{ckm} is important to check the consistency of the Standard Model 
and search for new physics. 
Various methods using $B\to D K$ decays have been introduced \cite{bdk_gamma}
to measure the unitarity triangle angle $\phi_3$ 
but these 
require knowledge of the branching fractions for different $B \to D K$
modes and therefore suffer from systematic errors due to detector response
and background contamination.

A novel technique based on the analysis of the three-body decay of 
the $D^0$ meson
\cite{dalitz_bdk} does not involve absolute branching 
fraction measurements and is, therefore, free from corresponding detector 
systematics. At the same time, the uncertainties in this
method due to the $D^0$ decay model can be controlled.

This method is based on two key observations:  neutral $D^{0}$ and $\bar{D}^{0}$
mesons can decay to a common final state such as $K_S \pi^+ \pi^-$, and the decay
$B \to D^{0} K^{(*)}$ can produce neutral $D^{0}$ mesons of both flavors via $b\to c\bar{u}s$
and $b\to u\bar{c}s$ transitions (see Figure 1), where the relative phase between the two interfering
amplitudes is the sum $\delta + \phi_3$ of strong and weak interaction
phases.  In the charge conjugate mode, the relative phase is the difference $\delta -
\phi_3$ of these phases, so both phases can be extracted from the measurements of such
$B$ decays and their charge conjugate modes. The annihilation diagram contribution is
neglected here since it affects only the extracted value of the strong phase $\delta$.
In our case, the measurement is based
on the analysis of Dalitz distribution of the three body final state of the 
$D^{(*)}$ meson. 

In this paper, we will discuss in detail the decay $B^+\to\bar{D^0}K^+$ state 
only, although a similar approach can be applied to other processes such as
$B^0\to D^0 K^{*0}$ or $B^{\pm}\to D^{*0} K^{\pm}$.

\section{Description of the method}

In the Wolfenstein parameterization of the CKM matrix, the amplitudes of the two diagrams
shown in Fig.~1 that contribute to the decay $B^+\to D^0(\bar{D}^0)K^+$
are given by $M_1\sim V_{cb}V_{us}^*\sim A\lambda^3$ and 
$M_2\sim V_{ub}V_{cs}^*\sim A\lambda^3(\rho+i\eta)$. 
These amplitudes interfere if the $D^0$ and $\bar{D}^0$ mesons decay
into the identical final state $K_s \pi^+ \pi^-$; 
we denote the admixed state as $\tilde{D}^0$. Assuming no CP 
asymmetry in $D^0$ decays, the amplitude of the $B^+$ decay can be written as
\begin{equation}
\label{intdist}
M_+=f(m^2_+, m^2_-)+ae^{i\phi_3+i\delta}f(m^2_-, m^2_+), 
\end{equation}
where $m_{+}^2$ and $m_{-}^2$ are the squared invariant masses of the $K_s \pi^+$ and
$K_s \pi^-$ combinations, respectively, and $f$ is the complex amplitude of the decay
$\bar{D}^0 \to K_S \pi^+ \pi^-$. The absolute value $a$ of the ratio between 
the two interfering amplitudes is given by

\[
a=\frac{|V_{ub}V_{cs}^*|}{|V_{cb}V_{us}^*|}\cdot\frac{|a_2|}{|a_1|} \sim
0.08\cdot 1.04/0.2196\cdot0.35, 
\]
where the color suppression factor $|a_2/a_1|$ can be estimated from the
Belle measurements of the color suppressed $\bar{B^0}\to D^0 \bar{K^0}$ 
\cite{bdk_color_supp} and color allowed $B^-\to D^0 K^-$ \cite{bdk_color_all}
decays:
\[
\frac{|a_2|}{|a_1|}=
\sqrt{\frac{Br(\bar{B^0}\to D^0 \bar{K^0})}
           {Br(B^-\to D^0 K^-)}}=0.35. 
\]
Though there are estimations of the $a$ value as large as 0.2 \cite{gronau}, 
in our Monte Carlo (MC) simulations, we have used the 
conservative value of $a=0.125$. 

Similarly, the amplitude of the charge conjugate $B^-$ decay is 
\[
M_-=f(m^2_-, m^2_+)+ae^{-i\phi_3+i\delta}f(m^2_+, m^2_-). 
\]
Once the functional form of $f$ is fixed by a choice of a model for $D^0 \to K_s \pi^+ \pi^-$
decays, the Dalitz distributions for $B^+$ and $B^-$ decays can be fitted simultaneously
by the above expressions for $M_{+}$ and $M_{-}$, with $a$, $\phi_3$, and $\delta$ as free
parameters.

Ref. \cite{dalitz_bdk} suggests a model-independent way of determining $\phi_3$ via 
a binned Dalitz plot analysis. However, this method is likely to 
introduce higher statistical errors in our present statistics-limited stage, and
certain technical issues involving backgrounds and reconstruction efficiency 
have to be solved. 
Here, we use a model-dependent approach with an unbinned maximum likelihood
fit of the Dalitz distribution, making optimal use of our small event sample.

\section{Study of sensitivity and model uncertainties}

The accuracy of the $\phi_3$ determination and possible systematic 
effects are determined from a MC simulation, where we generate
a large number of 
$\tilde{D^0}\to K_s\pi^+\pi^-$ decays according to a selected model
of the $D^0$ decay, 
then perform the unbinned likelihood fit of the Dalitz distribution
to obtain the parameter values and their errors.

The Dalitz plot for the decay $B^+\to [K_s\pi^+\pi^-]_{\bar{D^0}}K^+$ 
in the variables $m^2_+$ and $m^2_-$ was generated according to 
relation (\ref{intdist}), with the $D^0\to K_s\pi^+\pi^-$ decay amplitude 
$f$ described by the set of resonances listed in Table \ref{reslist}. 
The model used is similar to that in the CLEO 
measurement~\cite{dkpp_cleo} except that tensors and $K^*(1680)$ are excluded. 
The amplitude $f$ in this model is represented
by a coherent sum of $N$ two-body decay amplitudes plus one non-resonant decay amplitude:
\[
  f(m^2_+, m^2_-) = \sum\limits_{j=1}^{N} a_j e^{i\alpha_j}A_j(m^2_+, m^2_-)+
    b e^{i\beta}, 
\] 
where $N$ is the number of resonances, $A_j(m^2_+, m^2_-)$, $a_j$ and 
$\alpha_j$ are the matrix element, amplitude and phase, respectively, 
for the $j$-th resonance, and $b$ and $\beta$ are the amplitude
and phase for the non-resonant component. The total phase and amplitude 
are arbitrary. To be consistent with CLEO, we have chosen the $K_s\rho$
mode to have unit amplitude and zero relative phase. 
The description of the matrix elements $A_j$ follows Ref.~\cite{cleo_model}. 

The simulated Dalitz plot of the $\bar{D^0}$ decay is shown in 
Fig.~\ref{d0_plot}. The
corresponding distribution for the admixed $\tilde{D}^0$ decay contains the information about
the interference phase and amplitude between $D^0$ and $\bar{D}^0$ that can be extracted
from the fit to the Dalitz distribution.

Though in general the whole Dalitz plot 
is sensitive to the variations of the phase between $D^0$ and $\bar{D^0}$, 
the contribution from the interference is particularly high in certain regions. 
Since the maximum of the $\bar{D^0}$ plot density 
is in the
region of $K^*(892)^+$, the most sensitive area is the region of double 
Cabibbo suppressed $K^*(892)^-$, which 
interferes with the maximum amplitude of color suppressed $D^0$. Another 
sensitive part is the region of $K_s\rho$. Fig.~\ref{slices} demonstrates 
the CP asymmetry on the most sensitive regions of the Dalitz distribution.

For the estimation of the statistical error of $\phi_3$, samples 
of $10^4$ events were generated with total phase $\theta$ varying from $0$ to $2\pi$
($\theta=\delta+\phi_3$ for $B^+\to \bar{D^0} K^+$ and
$\theta=\delta-\phi_3$ for $B^-\to D^0 K^-$ decay) and 
the resulting $\tilde{D^0}$ distributions were fitted with $a$ and $\theta$
as free parameters. This is equivalent to the case 
where both $B^+$ and $B^-$ decays are fitted with ($a_+$, $\theta_+$)
and ($a_-$, $\theta_-$) as free parameters and the weak phase $\phi_3$
is extracted as
\[
\phi_3 = \frac{\theta_+ - \theta_-}{2}. 
\] 
The error of $\phi_3$ is slightly overestimated in this case since we
do not constrain the relative amplitudes $a_{+}$ and $a_{-}$ to be equal.

The error of $\phi_3$ was determined as 
\begin{equation}
\sigma^2_{\phi_3}(\phi_3, \delta)=
\frac{1}{4}(\sigma^2_{\theta}(\delta+\phi_3) + \sigma^2_{\theta}(\delta-\phi_3)), 
\label{phi3_stat}
\end{equation}
where $\sigma_{\theta}$ is the error of the parameter $\theta$ 
extracted from the fit. The corresponding statistical error $\sigma_{\phi_3}$ 
obtained with relation (\ref{phi3_stat}) ranges 
from $3.0^{\circ}$ to $3.7^{\circ}$ with a mean value of $3.4^{\circ}$ 
(averaged over all $\phi_3$ and $\delta$). 

We expect about $10^3$ $B^+\to [K_s\pi^+\pi^-]_{\bar{D^0}}K^+$ decays 
from a 1000 fb$^{-1}$ sample will result in a $10^{\circ}$ statistical 
accuracy in $\phi_3$. 
The statistics of $B$ decays collected at the present time by $B$-factories is 
therefore insufficient to make a $\phi_3$ measurement with a reasonable precision.
However, the 140 fb$^{-1}$ sample of Belle could provide the evidence of a CP asymmetry. 

We expect the model used for the $\bar{D^0}\to K_s\pi^+\pi^-$ decay 
to be one of the main sources of systematic errors of $\phi_3$. 
We estimate this contribution by fitting our simulated samples, 
described earlier, to five alternate models
for the decay amplitude $f(m_{+}^2, m_{-}^2)$ (see Table~\ref{syst}). 
It can be seen that while 
Blatt-Weisskopf form factors and $q^2$ resonance width dependence have
a rather small effect on the model error, the exclusion of
wide resonances and doubly Cabibbo suppressed $K^*(892)$ causes 
a large bias of $\phi_3$. We take the value of $10^{\circ}$ as a rough
estimate of the model uncertainty for the method. This value was obtained
by taking only the narrow resonances (including doubly Cabibbo suppressed
$K^*(892)$) and approximating the others with a flat non-resonant term. 
This is motivated by the fact that the amplitude of the narrow resonances is
well approximated with a Breit-Wigner shape, whereas wide 
states ($\Gamma\sim M$) contain substantial theoretical uncertainties. 

Though doubly Cabibbo suppressed $K^*(892)$ has a small amplitude
compared to other resonances (its fit fraction is 0.34\% \cite{bdk_cleo}), 
the error in its amplitude transforms into a large systematic error since
the whole area of the resonance interferes with the Cabibbo allowed 
$K^*(892)$ component of the color suppressed amplitude. We have investigated 
this influence by generating a sample of $10^4$ events according to
the model of Table~\ref{reslist} with subsequent fitting of this 
Dalitz plot with the same model where the amplitude of doubly 
Cabibbo suppressed $K^*(892)$ is changed. This analysis indicates that the 
statistical error of 0.02 on the 
$K^*(892)$ amplitude obtained by CLEO \cite{dkpp_cleo} 
transforms into approximately $3^{\circ}$ bias in $\phi_3$. The errors of other 
amplitudes are expected 
to contribute even less, since they lie mainly outside the maximum 
sensitivity region of the Dalitz distribution.

Another source of systematics comes from various kinds of background. We expect
three main background contributions to affect the determination of $\phi_3$:
\begin{itemize}
\item Background uniformly distributed over the Dalitz plot (combinatorial 
background), 
\item Background from $B^{\pm}\to D^0(\bar{D^0}) \pi^{\pm}$ events
with a pion misidentified as a kaon (pion background), 
\item Background from $B^{\pm}\to D^0(\bar{D^0}) K^{\pm}$ events
with a wrongly associated kaon of opposite sign (``wrong tag" background). 
\end{itemize}

To consider the worst case scenario, our MC samples were generated with one or another of
these backgrounds and then fit without any background component.  Neglected combinatorial
background at the level of 5\% results in an $8^{\circ}$ bias in $\phi_3$.  Neglected $D\pi$
background, even at the level of 20\%, does not induce an appreciable bias in $\phi_3$.
The contribution of wrongly associated kaons, which tag the flavor of $D^0$,
has a larger effect on systematics, since half of such kaons have the wrong 
sign and such events will be misinterpreted as decays of a $D$ meson
of opposite flavor. These $D^0$'s will distort the most 
sensitive region of the Dalitz plot. According to our simulation, the 
5\% admixture of wrong tag background introduces a $12^{\circ}$ bias of $\phi_3$. 

\section{Determination of $D^0\to K_s\pi\pi$ decay model}

For our MC simulation we have used the $D^0$ decay model measured
by CLEO based on their data sample of 5300 events \cite{dkpp_cleo}. However, 
the integrated luminosity collected by Belle can offer a much bigger sample 
of $D^0$ events. We have analyzed 78 fb$^{-1}$ of BELLE data looking
for decays of $D^*(2010)^+\to D^0\pi^+$, $D^0\to K_s\pi^+\pi^-$
(and its charge conjugate). The selection was based on the 
$D^0$ candidate invariant mass ($|M_{K_s\pi\pi}-M_{D^0}|<9$ MeV/c$^2$), 
and the mass difference of the $D^{*+}$ and $D^0$ candidates
($4.39<\Delta M<6.19$ MeV/c$^2$). We require the CM momentum of the
$D\pi$ system to be above 2.7 GeV/c to avoid contamination from 
combinatorics of $b\bar{b}$ decays.
We have selected 57800 events satisfying these requirements. The fit 
of the $M_D$ distribution yields a background fraction of 5.6\%. 

The Dalitz plot for the selected $D^* \to D\pi \to (K_S\pi\pi)\pi$ candidates
and its projections are shown 
in Fig.~\ref{ds2dpi_plot}. The data were fitted 
with a list of resonances from the CLEO model plus two scalars ($\sigma_1$ and
$\sigma_2$) in the $\pi\pi$ channel with free mass and width. 
We assumed the background and efficiency to be
uniform over the allowed phase space. 
Our fit result is shown in Table~\ref{fit_ds2dpi}. 

\section{Event selection}

The Belle detector has been described in detail elsewhere \cite{belle}. 
It is a large-solid-angle magnetic spectrometer consisting of a three-layer
silicon vertex detector (SVD), a 50-layer central drift chamber (CDC) for
charged particle tracking and specific ionization measurement ($dE/dx$), 
an array of aerogel threshold \v{C}erenkov counters (ACC), time-of-flight
scintillation counters (TOF), and an array of 8736 CsI(Tl) crystals for 
electromagnetic calorimetry (ECL) located inside a superconducting solenoid coil
that provides a 1.5 T magnetic field. An iron flux return located outside 
the coil is instrumented to detect $K_L$ mesons and identify muons (KLM).

Separation of kaons and pions is accomplished by combining the responses of 
the ACC and the TOF with $dE/dx$ measurement in the CDC to form a likelihood 
$\mathcal{L}(h)$ where $h$ is a pion or a kaon. Charged particles are 
identified as pions or kaons using the likelihood ratio (PID):
\[
  \mbox{PID}(K)  =\frac{\mathcal{L}(K)}{\mathcal{L}(K)+\mathcal{L}(\pi)}; 
  \mbox{PID}(\pi)=\frac{\mathcal{L}(\pi)}{\mathcal{L}(K)+\mathcal{L}(\pi)}=
  1-\mbox{PID}(K).
\]

The data sample of $152\cdot 10^6$ $B\bar{B}$ events (140 fb$^{-1}$) collected 
by the Belle detector was processed to search for $B^{\pm}\to D^0 K^{\pm}$
decays with $D^0$ decaying into the $K_s\pi^+\pi^-$ final state. In addition, 
we select $B^{\pm}\to D^0 \pi^{\pm}$ and $B^{\pm}\to D^*(2007)^0 \pi^{\pm}$
decays which serve as test samples for our fitting procedure and are used
for background analysis.

Neutral kaons are reconstructed from pairs of oppositely charged pions.
We require the reconstructed vertex distance from the interaction point 
in the plane transverse to the beam axis to be more than 1 mm 
and the invariant mass $M_{\pi\pi}$
to be $|M_{\pi\pi}-M_{K_s}|<10$ MeV/c$^2$. 

Selection of $B$ candidates is based on the CM energy difference
$\Delta E = \sum E_i - E_{beam}$ and $B$ meson beam-constrained mass
$M_{bc} = \sqrt{E_{beam}^2 - (\sum p_i)^2}$, where $E_{beam}=m_{\Upsilon}/2$ is the CM beam 
energy, and $E_i$ and $p_i$ are the CM energies and momenta of the
$B$ candidate decay products. We select events with $M_{bc}>5.2$ GeV/c$^2$
and $|\Delta E|<0.2$ GeV for the analysis. The cuts for signal candidates are
$5.272<M_{bc}<5.288$ GeV/c$^2$ and $|\Delta E|<0.022$ GeV. 
In addition, we apply a cut on the invariant mass of the $D^0$ candidate: 
$|M_{K_s\pi\pi}-m_D|<11$ MeV/c$^2$. 

To suppress the continuum background, we require $|\cos\theta_{thr}|<0.8$, 
where $\theta_{thr}$ is the angle between the thrust axis of the $B$ candidate 
and the rest of the event. For additional background rejection, we 
use the Fisher discriminant composed of 11 parameters \cite{fisher}: 
the production angle of the $B$ candidate, the angle of the $B$ thrust axis relative
to the beam axis and nine parameters representing the momentum flow in the event
relative to the $B$ thrust axis in the CM frame.

The selection efficiency for these cuts is 12\%, based on MC simulation, 
implying an expectation of $158 \pm 33$ signal events in our data sample.
The distributions for 
$\Delta E$, $M_{bc}$ and $M_D$ for the experimental data are shown in 
Fig.~\ref{b2dk}. The number of $B\to D^0 K$ events satisfying all selection criteria is 138.

To fit the $M_{bc}$ distribution, we use the empirical background function 
suggested by the ARGUS collaboration \cite{argus} plus a Gaussian signal. 
The $\Delta E$ distribution is fit with a linear background plus two Gaussians
representing the signal (with a mean value around zero) and background
from $B\to D^0\pi$ with a pion misidentified as a kaon (with 
$\Delta E\sim$50 MeV). For the $M_D$ fit we use a constant background plus 
a combination of two Gaussians for the signal.
The resolutions of the parameters were extracted from the fit of 
$B\to D^0\pi$ data, described below.

The fit of the $\Delta E$ distribution yields $107\pm 12$ signal events and
$33\pm 3$ background events, consistent with the expectation of $158 \pm 33$ signal events.
The Dalitz plots of the selected candidates and their projections 
onto $M^2_{K_s\pi^+}$, $M^2_{K_s\pi^-}$ and $M^2_{\pi^+\pi^-}$ axes 
are shown separately for $B^-$ and $B^+$
decays in Fig.~\ref{b2dk_plot_m} and Fig.~\ref{b2dk_plot_p},
respectively. Note that for $B^+$ decays the axes are chosen so that 
the $K^*$ band is aligned along the $y$-axis, as for $B^-$ decays. 

The $B^{\pm}\to D^0 \pi^{\pm}$ selection cuts are the same as those for 
$B\to D^0 K$, except that the cut on $K/\pi$ PID probability is modified to 
select a pion ($Pr(K/\pi) > 0.5$). 
The same fit functions are used to fit $\Delta E$, $M_{bc}$ and $M_D$ distributions
as for $B\to D^0 K$ (with only one Gaussian peak in $\Delta E$), and
the fitted distributions are shown in Fig.~\ref{b2dpi}.
The extracted resolutions of the selection parameters, used in the $B \to D^0K$ 
fits, are 13 MeV for $\Delta E$, 
3.2 MeV/c$^2$ for $M_{bc}$ and $6.3$ MeV/c$^2$ for $M_D$. 
The Dalitz distribution of the $D^0\to K_s\pi^+\pi^-$ decay from $B\to D^0\pi$ process
is shown in Fig.~\ref{b2dpi_plot}. 

\section{Backgrounds}

To fit the Dalitz plot correctly, backgrounds have to be taken into 
account in the fit model. For each of the background contributions, 
we need to obtain not only its fraction in the event sample, but also 
the Dalitz plot shape. 

The largest contribution for the decay of our interest comes from the
continuum $q\bar{q}$ background ($q=u, d, s, c$). 
It includes the background with purely 
combinatorial tracks, and continuum $D^0$ mesons combined with a 
random kaon. This type of background is analyzed using the off-resonant 
statistics collected below $\Upsilon(4S)$ as
well as on-resonance data with $\cos\theta_{thr}$ and Fisher cuts
applied to select continuum. 

The background coming from $b\bar{b}$ events originates either from 
$B\to D^0 K$ decay with some final state particles interchanged with
combinatorial ones from the decay of the other $B$ meson, or from other
charged or neutral $B$ decays (possibly with misidentified or lost 
particles). We subdivide the $b\bar{b}$ background into four types. 

The $B\to D^0\pi$ process with a pion misidentified as a kaon is suppressed by the
requirement on the $K/\pi$ identification probability and CM energy difference. 
The fraction of this background is obtained by fitting the $\Delta E$ 
distribution; the Dalitz plot distribution is that of $D^0$ without 
the opposite flavor admixture. 

Other decays of charged and neutral $B$ mesons contributing to the 
background are investigated with a generic MC sample. Most surviving candidates 
are due to decay of one of the $B$ mesons to the $D^{*0} X$ state, 
with some particles taken from the other $B$ decay. 

Events where one of the $D^0$ decay products is substituted by
a combinatorial kaon or pion were studied using the MC data set where 
one of the $B$ mesons from the $\Upsilon(4S)$ decays into the $D^0 K$ state. 

Events where a valid $D^0$ is combined with a random flavor-tagging kaon 
are of importance, because half of the kaons have 
the wrong sign: such events will be misinterpreted 
as decays of $\bar{D^0}$, thus introducing distortion in the
most sensitive area of the Dalitz plot. This background was studied
with the same MC sample as for $D^0$ combinatorics. No events 
of this kind were found, which allows us to put an upper limit of 0.4\%
(at 95\% C. L.) on the fraction of this contribution. 

The Dalitz plot shapes of all the backgrounds were parameterized with 
empirical functions that included polynomial terms as well as Breit-Wigner
peaks. The investigation of the $\Delta E$ distribution for all considered 
backgrounds allows us to conclude that none of them except combinatorics in 
$D^0$ decay is peaked in $\Delta E$. Therefore, a $\Delta E$ fit gives 
a good estimate of the total background rate, which is 23\%. 
The fractions of each contribution are as follows: 
14\% for continuum combinatorics, 3\% for $D^0$ from continuum, 
0.5\% for $B\to D^0\pi$, 4\% for decays of $B^{\pm}$ other than 
$B\to D^0 K/\pi$, 2\% for decays of neutral $B$ and 0.4\% for combinatorics
in $D^0$ decay. 

Another approach to take into account backgrounds is to use events from
$D^0$ sidebands. Since all the backgrounds except $D^0$ from
continuum and $D^0$ decay combinatorics are not peaked in the $D^0$ invariant
mass distribution, the $D^0$ sidebands are a good approximation of the 
total background shape. The $D^0$ combinatoric background has a small 
fraction and can be
neglected, and the shape of the continuum $D^0$ background is known and can be added
separately. However, the statistics in $B\to D^0 K$ decays is not sufficient
to fit the background shape with good precision, so we use the $B\to
D^0\pi$ sample for this purpose. 
This approach is used together with the flat background distribution to 
estimate the systematic uncertainty caused by our background parameterization. 

\section{Fit of $D^0$ Dalitz plot}

We use the unbinned likelihood technique to fit the Dalitz plot of both 
MC simulated and experimental samples. In the fit
function of experimental data, finite momentum resolution and efficiency 
were taken into account. 

The value which is minimized in the fit is the inverse logarithm of the 
unbinned likelihood function
\[
  -2 \log L = -2\left[\sum\limits^n_{i=1}\log p(m^2_{i+}, m^2_{i-}) - 
  \log\int\limits_D p(m^2_+, m^2_-)dm^2_+ d m^2_-\right], 
\]
where $m^2_{i+}$, $m^2_{i-}$ are measured (or generated) Dalitz plot
variables and $p(m^2_+, m^2_-)$ is the Dalitz plot density being fitted. 
The Dalitz plot density is a sum of the signal $D^0-\bar{D^0}$ interference 
$|f(m^2_+, m^2_-)+ae^{-i\phi}f(m^2_-, m^2_+)|^2$ and the background distribution, 
convoluted with
a momentum resolution function and multiplied by the efficiency shape. 
The efficiency and resolution functions were obtained with a MC simulation
with $D^0$ decaying according to the phase space distribution. 

To check the consistency of the $\phi_3$ fit, the same fitting procedure was 
applied to $B\to D^{(*)0}\pi$ and $D^{*}\to D^0\pi$ test samples as well as 
the $B\to D^0 K$
signal. The idea is to check the absence of opposite flavor amplitude $a$
that can appear if the Dalitz plot shape is not described well by the fit model. 
In the case of $D^{*}\to D^0\pi$ data, we do not expect any contribution 
from opposite flavor (provided flavor tagging is correct), but this sample 
cannot serve as a good fit test since the same data are used to fit the $D^0$
decay model. So, these data can only indicate the defects of $D^0$ decay
parameterization, which can affect the $\phi_3$ fit. The sample of $D^0$ from 
$B\to D^{(*)0}\pi$ decay allows for a more reliable test of the fit procedure, since 
it is independent of the sample used to determine the model, and the background
sources are expected to be similar to those for our signal. 
In the case of $B\to D^{(*)0}\pi$, however, a small opposite flavor amplitude is expected
($\sim |V_{ub} V^*_{cd}|/|V_{cb}V^*_{ud}|$) with $a\sim 0.01-0.02$. 

For our $\phi_3$ fit check, the free parameters for Dalitz plot fit were the
relative amplitude $a$ and total phase $\theta$. 

In the $D^{*}\to D^0\pi$ data fit, we have used the flat background with the same 
fraction as used in the case of determination of the $D^0$ model. 
The result of the fit is the following:
$a=0.0041\pm 0.0050$, $\theta=123^{\circ}\pm 62^{\circ}$ for $D^0$ decay and 
$a=0.0047\pm 0.0050$, $\theta=114^{\circ}\pm 52^{\circ}$ for $\bar{D^0}$ decay, which is 
consistent with zero.

The fit results for $B\to D^0\pi$ are $a_-=0.067\pm 0.027$, $\theta_-=228^{\circ}\pm 23^{\circ}$ for 
$D^0$ from $B^-\to D^0\pi^-$ decay and $a_+=0.065\pm 0.029$, $\theta_+=232^{\circ}\pm 24^{\circ}$
for $\bar{D^0}$ from $B^+\to D^0\pi^+$ decay. It should be noted that since the value 
of $a$ is positive definite, the error of this parameter does not serve as a good 
measure of the $a=0$ hypothesis. To demonstrate the deviation of the opposite 
flavor amplitude from zero, the real and imaginary parts of the complex 
relative amplitude $a\exp(-i\theta)$ are more suitable. 
Fig.~\ref{constr_plot} shows complex relative
amplitude constraints for $B^-$ and $B^+$ data separately. 
It can be seen from the plot that both amplitudes differ from the expected value by 
more than two standard deviations.

As another check of the consistency of our fit result with the $a=0$ hypothesis, we have
performed a toy MC study. A Dalitz plot of 1700 events (which corresponds
to full statistics of $B\to D^0\pi$ with both flavors included) was generated 
in 566 trials with the model where $a=0$, and the fit of these data was performed 
with $a$ and $\phi$ allowed to float. The distribution of the parameter $a$ 
for these trials is shown in Fig.~\ref{a_mc}. It can be seen from the plot
that only $\sim 1$\% of the $a$ values are above 0.06. The bias observed 
needs further investigation. One of the possible reasons can be the variation 
of the background shape that is not parameterized by the fit function. 
Also, one cannot discount statistical fluctuation. 

Another test sample of $D^0$'s with the small opposite flavor admixture
is provided by the $B\to D^{*0}\pi$ decay with $D^{*0}$ decaying into
$D^{0}\pi^0$. The constraints on the complex amplitude are shown in 
Fig.~\ref{constr_plot} (right). In this case, we do not
observe any significant opposite flavor amplitude. The result of the fit is 
$a_-=0.057\pm 0.054$, $\theta_-=340\pm 65$ for $D^0$ from $B^-\to D^{*0}\pi^-$ decay and 
$a_+=0.041\pm 0.069$, $\theta_+=164\pm 100$ for $\bar{D^0}$ from $B^+\to D^{*0}\pi^+$.
Though this sample is smaller (350 events) and it can not serve
a reliable test of fit bias, we observe phases different from $B\to D^0\pi$ case, 
which can indicate that the fitting procedure itself does not introduce a 
bias to the phase.

For $D^0-\bar{D^0}$ interference fit in $B\to D^0 K$, we have performed a simultaneous
fit with free parameters $a$, $\phi_3$ and $\delta$. 
The result of the fit overlaid on the experimental Dalitz distribution 
projections in shown in Fig.~\ref{b2dk_plot_m} and Fig.~\ref{b2dk_plot_p} for
$B^-$ and $B^+$ decays, respectively. 
The fit yields $a=0.33\pm 0.10$, $\phi_3=92^{\circ}\;^{+19}_{-17}$, 
$\delta=165^{\circ}\;^{+17}_{-19}$.
It should be noted that the fit errors are quite non-parabolic, since 
the phase error obtained from the fit is inversely proportional to the value of 
$a$. Therefore, the errors quoted can not reliably represent the accuracy of the
measurement. The constraint plots on the pairs of parameters ($\phi_3$,
$\delta$)
and ($a$, $\phi_3$) shown in Fig.~\ref{b2dk_free} give more information about the
accuracy and the statistical significance of the fit. 
The plots show one, two and three standard deviation contours of each pair
of parameters. The central values are marked with crosses. 
For each of the plots, the third parameter was allowed to float. 

Note the two-fold ambiguity of this method: ($\phi_3, \delta$) and 
($\phi_3+\pi, \delta+\pi$) cannot be distinguished, as they lead
to the same total phases. Since there are no reliable measurements or 
theoretical predictions of the value of the strong phase, we keep
both values as solutions for our data. 

The $\phi_3$ statistical significance $-2\log L/L_{max}$ is presented in 
Fig.~\ref{b2dk_asym} (left). The $a$ and $\delta$ parameters were allowed to 
float. 

To illustrate the observation of CP asymmetry and its significance, we have 
performed a separate fit of the $B^+$ and $B^-$ data sets. The results of the
fits are shown in Fig.~\ref{b2dk_asym} (right) as constraints on the complex
relative amplitude $a\exp(i\theta)$, as in the case of test sample fits. 
The relative amplitude of the $B\to D^0\pi$ test sample sample fit is also 
shown for comparison. The asymmetry of three standard deviations is apparent. 

The influence of the different sources of systematics on our fit result are 
presented in Table~\ref{syst_table}. The systematic error is 
dominated by the $D^0$ model uncertainty. The component related to the 
background parameterization was estimated by extracting the background shape 
from the $M_D$ sidebands and by using the flat background distribution. 
As mentioned above, the efficiency shape and resolution were extracted from 
MC simulation. To estimate their contribution to the systematic error, we 
have used the flat efficiency and ``ideal" resolution, respectively. 

The nonzero opposite flavor amplitude we observe in the $B\to D^0\pi$ test sample
may indicate some systematic effect such as background structure or a deficiency 
of the $D^0$ decay model. Since the source of this bias is indeterminate, 
we conservatively treat it as an additional systematic effect. 

Since the relative amplitude $a$ is positive definite, its fit value is biased 
towards the higher values by approximately one standard deviation. 
Consequently, the strong dependence of 
the $\phi_3$ error on $a$ can cause an overestimation of the $\phi_3$
determination accuracy. However, the constraint plot of $a$ vs. $\phi_3$ 
(Fig.~\ref{b2dk_free}, right) shows no significant correlation between the parameters. 
Therefore, we have calculated the $\phi_3$ and $\delta$ for 
the value of $a=0.23$, which is one standard deviation lower than the value
obtained in the fit. 
With model and systematic errors taken into account, and 
$a$ bias correction applied, our fit result can be presented as:
\begin{itemize}
  \item Weak phase $\phi_3=95^{\circ}\;^{+25}_{-20}\pm 13^{\circ}\pm 10^{\circ}$, 
  \item Strong phase $\delta=162^{\circ}\;^{+20}_{-25}\pm 12^{\circ}\pm 24^{\circ}$. 
\end{itemize}
The first error is statistical, the second is experimental systematics and
the third is model uncertainty. 
Fig.~\ref{b2dk_90cl} shows 90\% CL regions for the pairs of parameters
($\phi_3$, $\delta$) and ($\phi_3$, $a$) with systematic uncertainty 
included. The 90\% confidence intervals for 
the fit parameters (including systematic uncertainty) are: 
$0.15<a<0.50$, $61^{\circ}<\phi_3<142^{\circ}$
and $104^{\circ}<\delta<214^{\circ}$. 

\section{Conclusion}

We have applied a new method to measure the CP-sensitive angle $\phi_3$ 
using the Dalitz analysis of the three-body $D^0$ decay in the process $B\to D^0 K$. 
The method is directly sensitive to the value of $\phi_3$ and is free from 
discrete ambiguities (for the $\phi_3$ values in the range from 0 to $180^{\circ}$). 
The statistical and systematic uncertainties of this method were studied
using MC simulation. The statistics of $1000$ fb$^{-1}$ should permit us to
determine $\phi_3$ with a $10^{\circ}$ statistical uncertainty.
We estimate the
model uncertainty of the method to be of the order of 10$^{\circ}$. This uncertainty
can be significantly reduced with the increase of associated experimental data, 
and practically eliminated by using the data from the $c\tau$-factory
\cite{dalitz_bdk}. 

The first measurement of $\phi_3$ using this technique
was performed based on 140 fb$^{-1}$ statistics collected by the Belle detector. 
The significance of the direct CP violation effect (including systematic
effects) is 2.4 standard deviations. The values of $\phi_3$ and relative 
amplitude $a$ obtained are in agreement with the SM expectation. 
The 90\% confidence interval obtained for the $\phi_3$ value is 
$61^{\circ}<\phi_3<142^{\circ}$. 
The accuracy 
of the measurement can be improved in the future by adding other 
suitable decay modes into the analysis. 

\section{Acknowledgements}
We wish to thank the KEKB accelerator group for the excellent
operation of the KEKB accelerator.
We acknowledge support from the Ministry of Education,
Culture, Sports, Science, and Technology of Japan
and the Japan Society for the Promotion of Science;
the Australian Research Council
and the Australian Department of Education, Science and Training;
the National Science Foundation of China under contract No.~10175071;
the Department of Science and Technology of India;
the BK21 program of the Ministry of Education of Korea
and the CHEP SRC program of the Korea Science and Engineering Foundation;
the Polish State Committee for Scientific Research
under contract No.~2P03B 01324;
the Ministry of Science and Technology of the Russian Federation;
the Ministry of Education, Science and Sport of the Republic of Slovenia;
the National Science Council and the Ministry of Education of Taiwan;
and the U.S.\ Department of Energy.

\clearpage

\begin{table}
\begin{center}
\caption{List of resonances used for $\bar{D^0}\to K_s\pi^+\pi^-$ decay
simulation.}
\vspace{\baselineskip}
\label{reslist}
\begin{tabular}{|l|c|c|} \hline
Resonance                    & Amplitude & Phase, (deg)  \\ \hline
$K^*(892)^+\pi^-$            & 1.56         & 150        \\ 
$K_s\rho^0$                  & $\equiv 1$   & $\equiv 0$ \\
$K^*(892)^-\pi^+$            & 0.11         & 321        \\
$K_s\omega$                  & 0.037        & 114        \\
$K_s f_0(980)$               & 0.34         & 188        \\
$K_s f_0(1370)$              & 1.8          & 85         \\
$K_0^*(1430)^+\pi^-$         & 2.0          & 3          \\
$K_s\pi^+\pi^-$ non-resonant & 1.1          & 160        \\ \hline
\end{tabular}
\end{center}
\end{table}

\begin{table}
\begin{center}
\caption{Estimation of $\phi_3$ model systematics.}
\vspace{0.5\baselineskip}
\label{syst}
\begin{tabular}{|c|c|} \hline
Fit model & $(\Delta\phi_3)_{max}$ (deg)\\
\hline
$K^*(892)$, $\rho$, nonres                              & 29.3 \\
$K^*(892)$, $\rho$, DCS $K^*(892)$, nonres              & 6.7  \\
$K^*(892)$, $\rho$, DCS $K^*(892)$, $f^0(980)$, nonres  & 9.9  \\
$F_r=F_D=1$                                             & 3.1  \\
$\Gamma(q^2)=Const$                                     & 4.7  \\
\hline
\end{tabular}
\end{center}
\end{table}

\begin{table}
\begin{center}
\caption{Fit results for $D^0\to K_s\pi^+\pi^-$ decay. Errors are statistical only.}

\vspace{0.5\baselineskip}
\label{fit_ds2dpi}
\begin{tabular}{|l|c|c|} \hline
Resonance                    & Amplitude   & Phase, (deg)\\ \hline
$K^*(892)^-\pi^+$            & $1.706\pm 0.015$
                             & $138.0\pm 0.9$ 
                             \\ 

$K_s\rho^0$                  & $1.0$ (fixed)                                 
                             & 0 (fixed)   
                             \\

$K^*(892)^+\pi^-$            & $(13.6\pm 0.8)\times 10^{-2}$
                             & $330\pm 3$
                             \\

$K_s\omega$                  & $(32.8\pm 1.8)\times 10^{-3}$
                             & $114\pm 3$
                             \\

$K_s f_0(980)$               & $0.385\pm 0.011$
                             & $214.2\pm 2.3$
                             \\

$K_s f_0(1370)$              & $0.49\pm 0.04$ 
                             & $311\pm 6$
                             \\

$K_s f_2(1270)$              & $1.66\pm 0.05$
                             & $341.3\pm 2.3$
                             \\

$K_0^*(1430)^-\pi^+$         & $2.09\pm 0.05$
                             & $353.6\pm 1.8$
                             \\

$K_2^*(1430)^-\pi^+$         & $1.20\pm 0.05$
                             & $316.9\pm 2.1$
                             \\

$K^*(1680)^-\pi^+$           & $1.62\pm 0.24$
                             & $84\pm 10$
                             \\

$K_s \sigma_1$ ($M_{\sigma_1}=535\pm 6$ MeV, $\Gamma_{\sigma_1}=460\pm 15$ MeV)
			     & $1.66\pm 0.09$
                             & $217.3\pm 1.4$
                             \\

$K_s \sigma_2$ ($M_{\sigma_2}=1063\pm 7$ MeV, $\Gamma_{\sigma_2}=101\pm 12$ MeV)
	                     & $0.31\pm 0.04$
                             & $257\pm 11$
                             \\

non-resonant                 & $6.51\pm 0.22$ 
                             & $149.0\pm 1.6$ 
                             \\ 

\hline
\end{tabular}
\end{center}
\end{table}

\begin{table}
\caption{Estimation of systematic uncertainties.}
\label{syst_table}
\begin{tabular}{|l|c|c|c|} \hline
Source               & $\Delta a$ & $\Delta\phi_3$, (deg) & $\Delta\delta$, (deg) \\ \hline
$D^0$ decay model    & 0.03       & 10                   & 24                   \\
Background shape     & 0.015      & 3.2                  & 2.3                  \\
Efficiency shape     & 0.010      & 2.4                  & 0.8                  \\
Momentum resolution  & 0.010      & 2.5                  & 0.6                  \\
Test sample bias     & 0.02       & 12                   & 12                   \\ \hline 
Total                & 0.04       & 16                   & 27                   \\ \hline

\end{tabular}
\end{table}

\clearpage

\begin{figure}
  \begin{center}
  \epsfig{figure=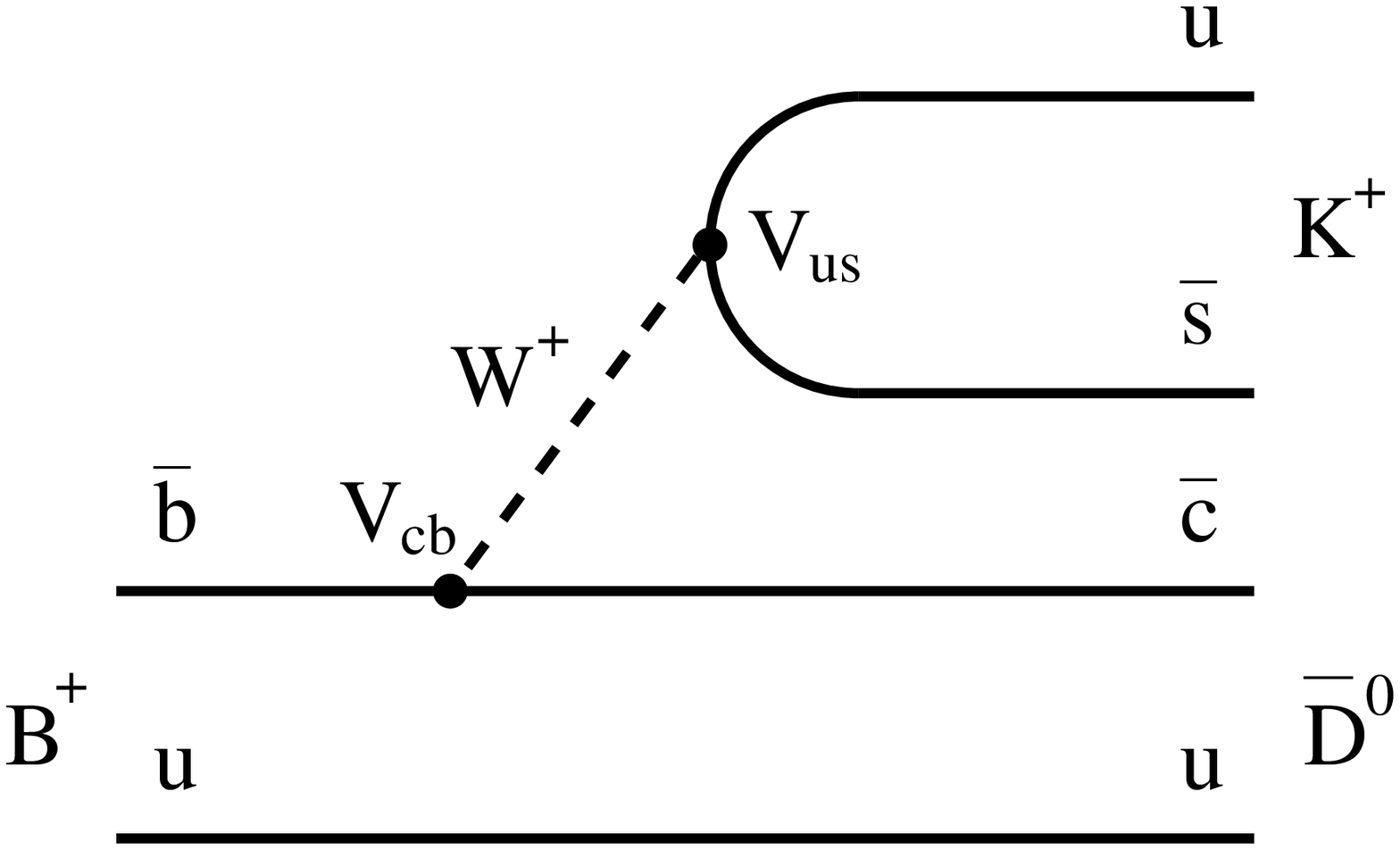,width=0.4\textwidth}
  \hspace{0.1\textwidth}
  \epsfig{figure=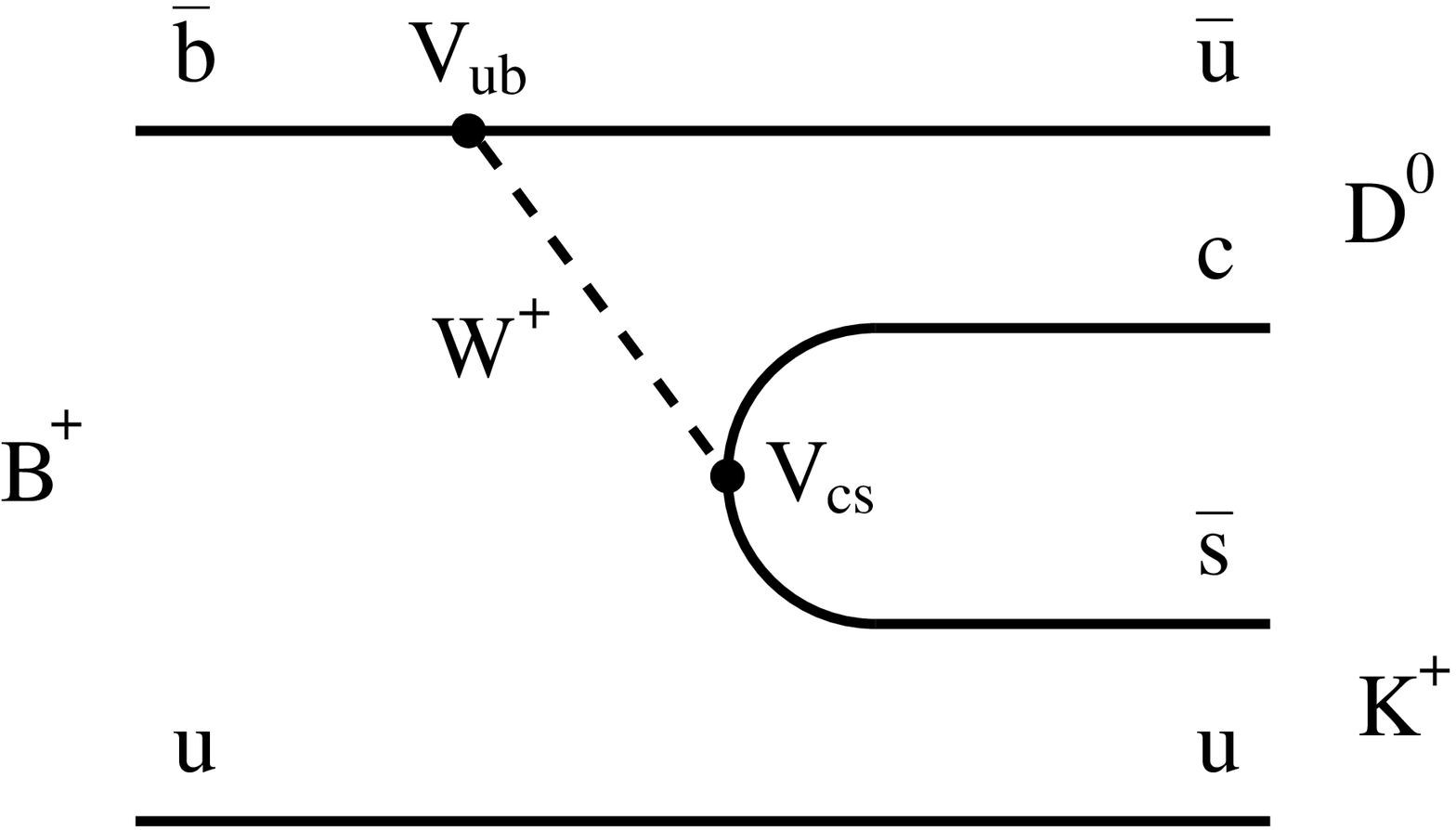,width=0.4\textwidth}
  \caption{Diagrams of $B^+\to \bar{D^0}K^+$ (left) and $B^+\to D^0K^+$ 
           (right) decays.}
  \label{diags}
  \end{center}
\end{figure}

\begin{figure}
  \epsfig{figure=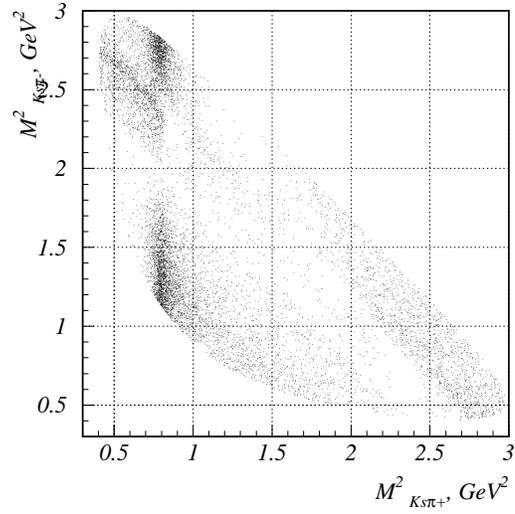,width=0.47\textwidth} 
  \caption{Simulated Dalitz plot of $\bar{D^0}\to K_s\pi^+\pi^-$ decay.}
  \label{d0_plot}
\end{figure}

\begin{figure}
  \vspace{-0.05\textwidth}
  \begin{center}
  \epsfig{figure=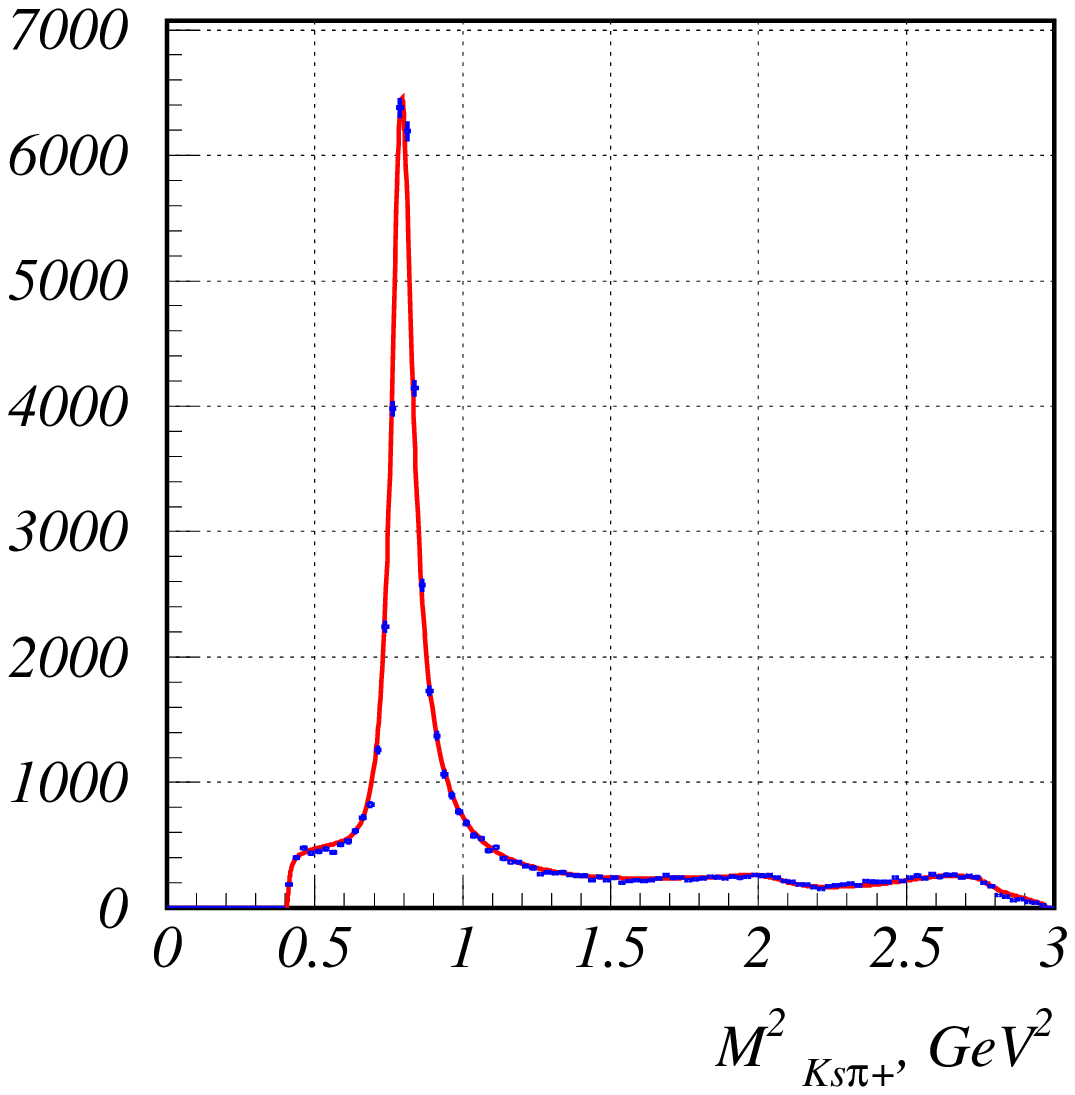,width=0.45\textwidth}
  \hspace{-0.05\textwidth}
  \epsfig{figure=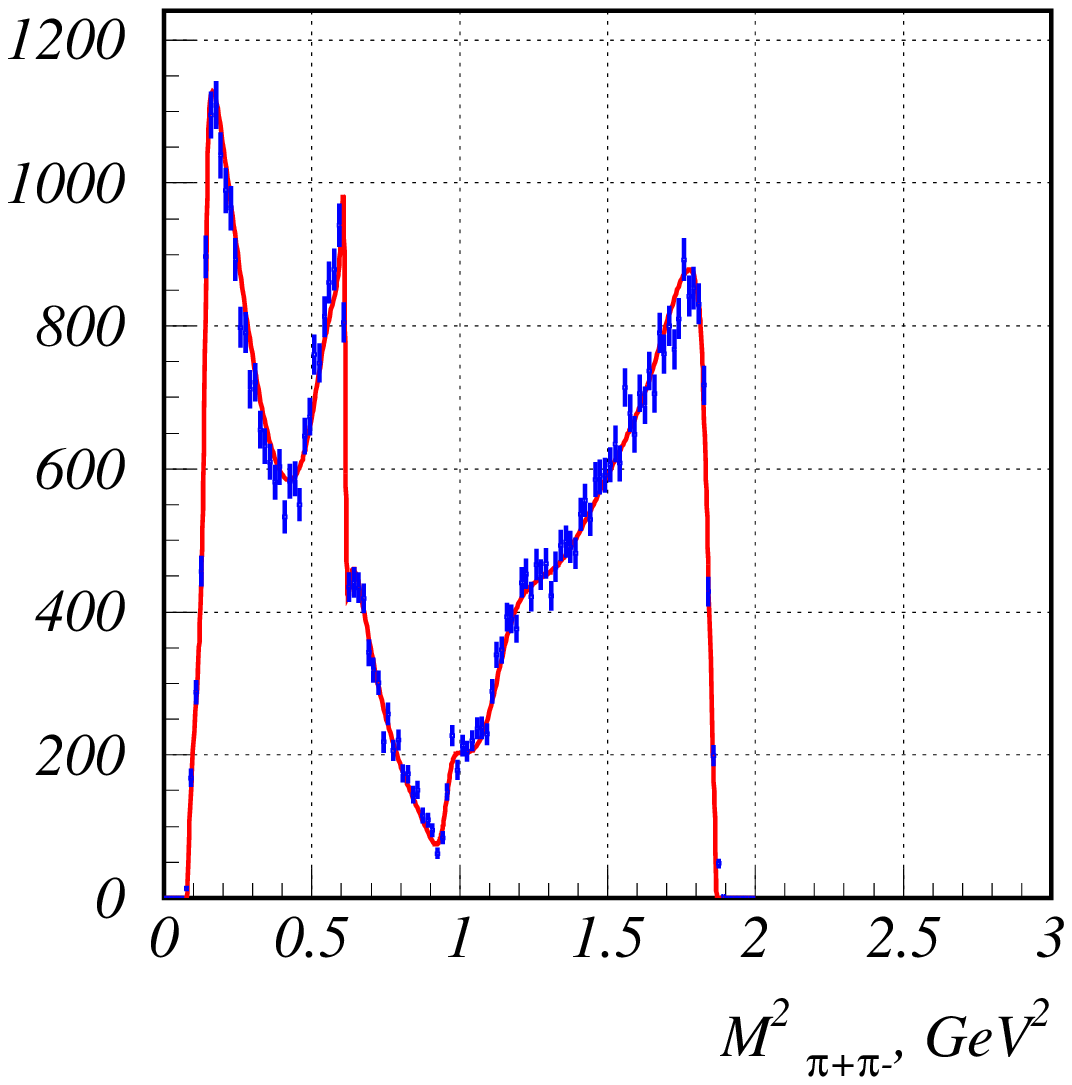,width=0.45\textwidth}
  \epsfig{figure=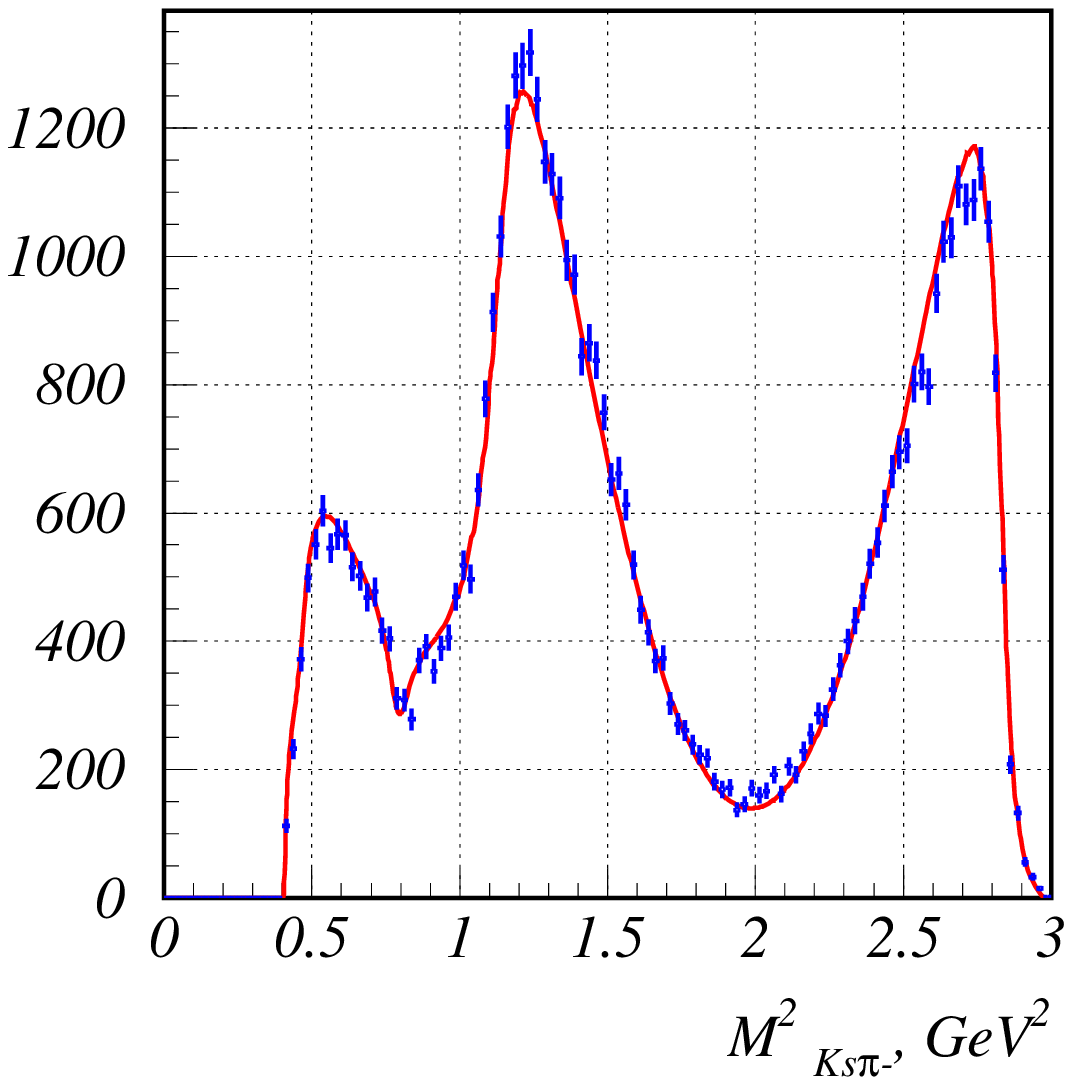,width=0.45\textwidth}
  \hspace{-0.05\textwidth}
  \epsfig{figure=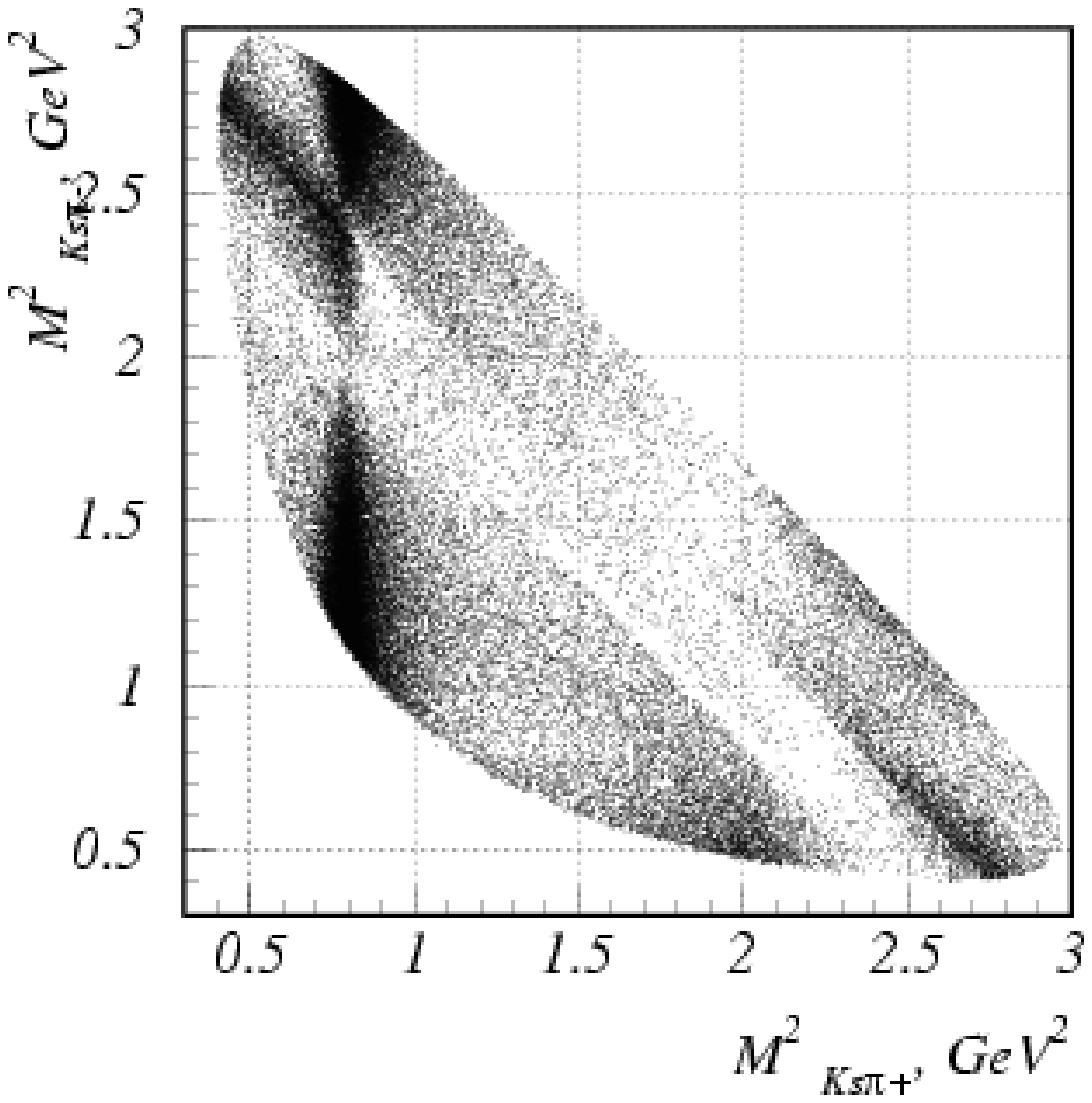,width=0.45\textwidth}
  \caption{$M^2_{K\pi^+}$, $M^2_{K\pi^-}$, $M^2_{\pi^+\pi^-}$ distributions
   and Dalitz plot of $D^0\to K_s\pi^+\pi^-$ decay from $D^{*+}\to D^0\pi$
   process.}
  \label{ds2dpi_plot}
  \end{center}
\end{figure}

\begin{figure}
  \vspace{-0.05\textwidth}
  \begin{center}
  \epsfig{figure=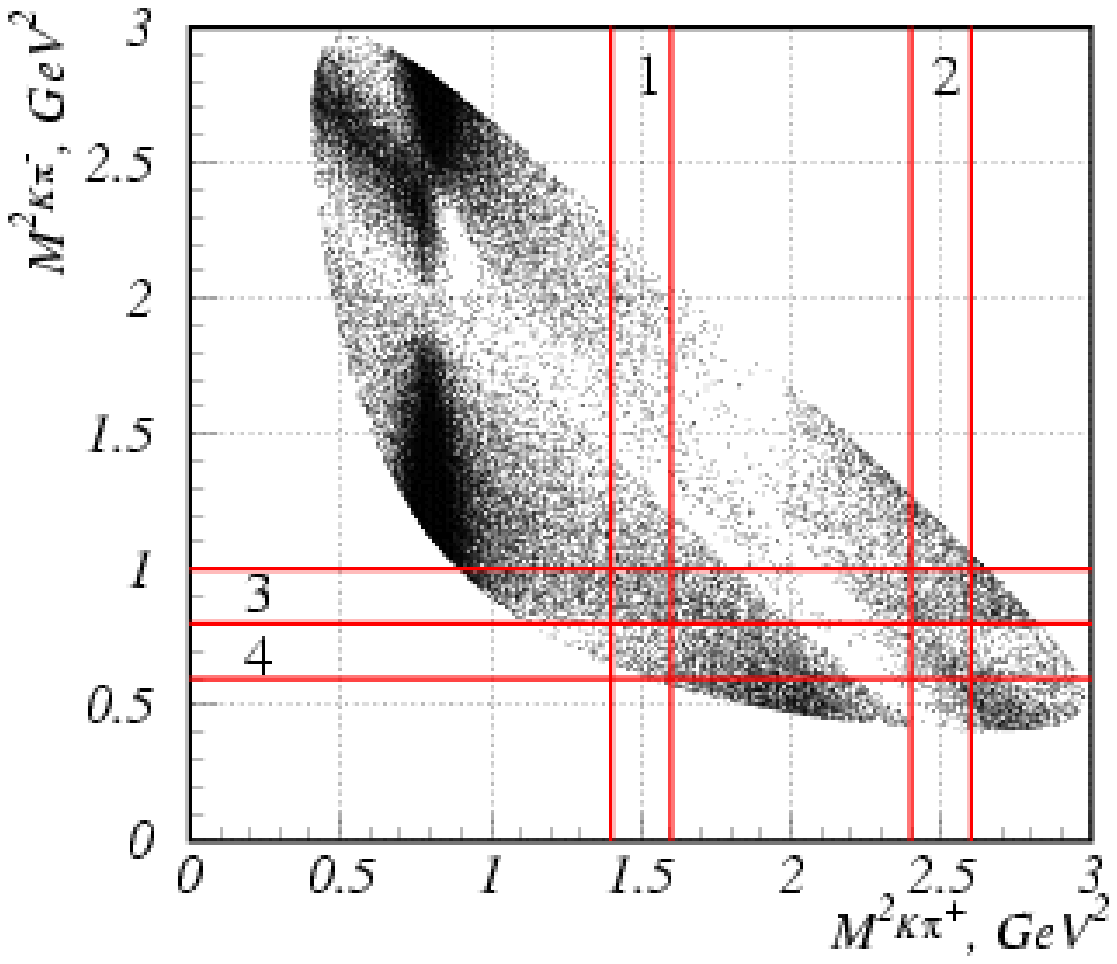,width=0.45\textwidth}

  \vspace{-0.05\textwidth}
  \epsfig{figure=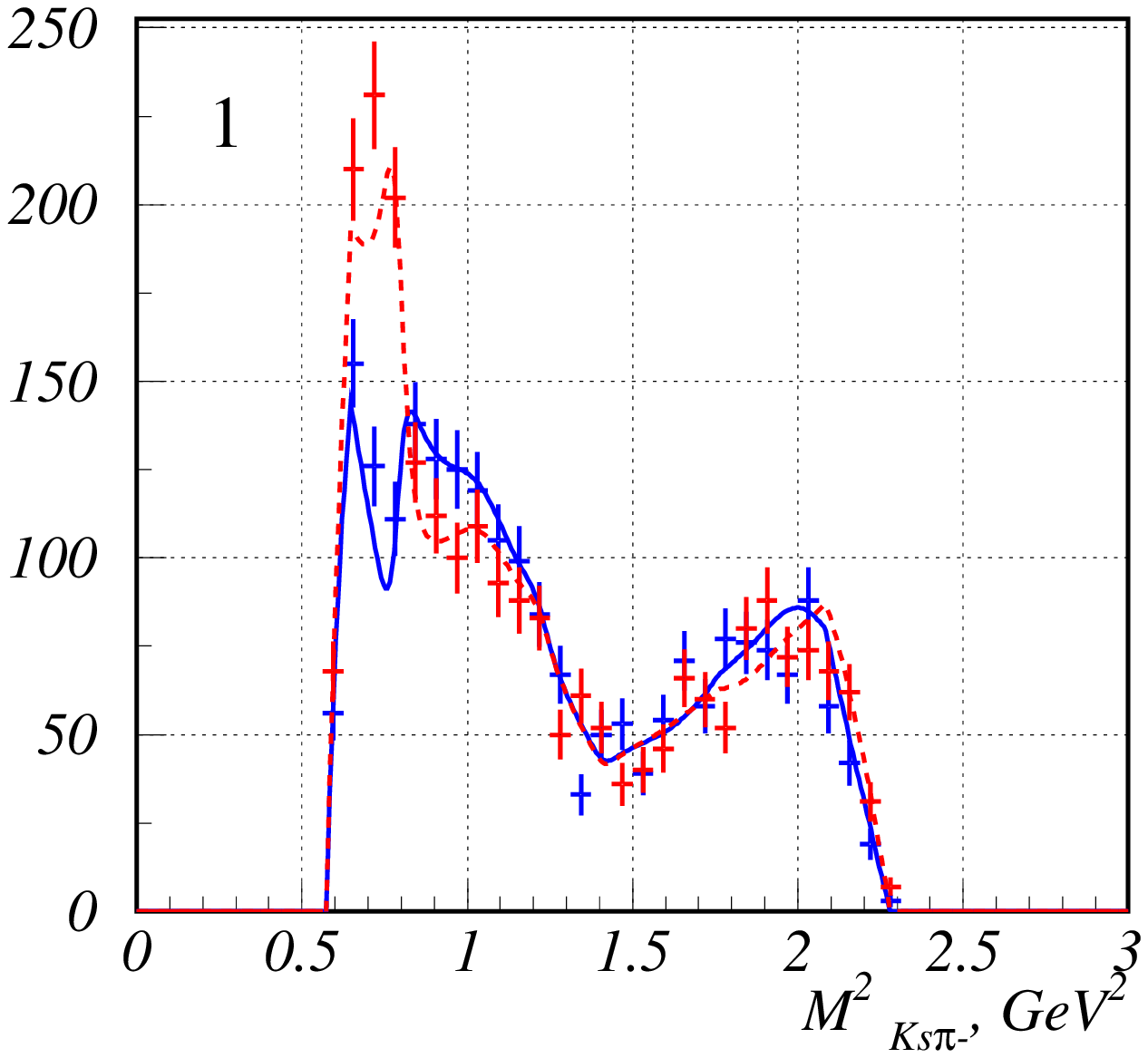,width=0.45\textwidth}
  \hspace{-0.05\textwidth}
  \epsfig{figure=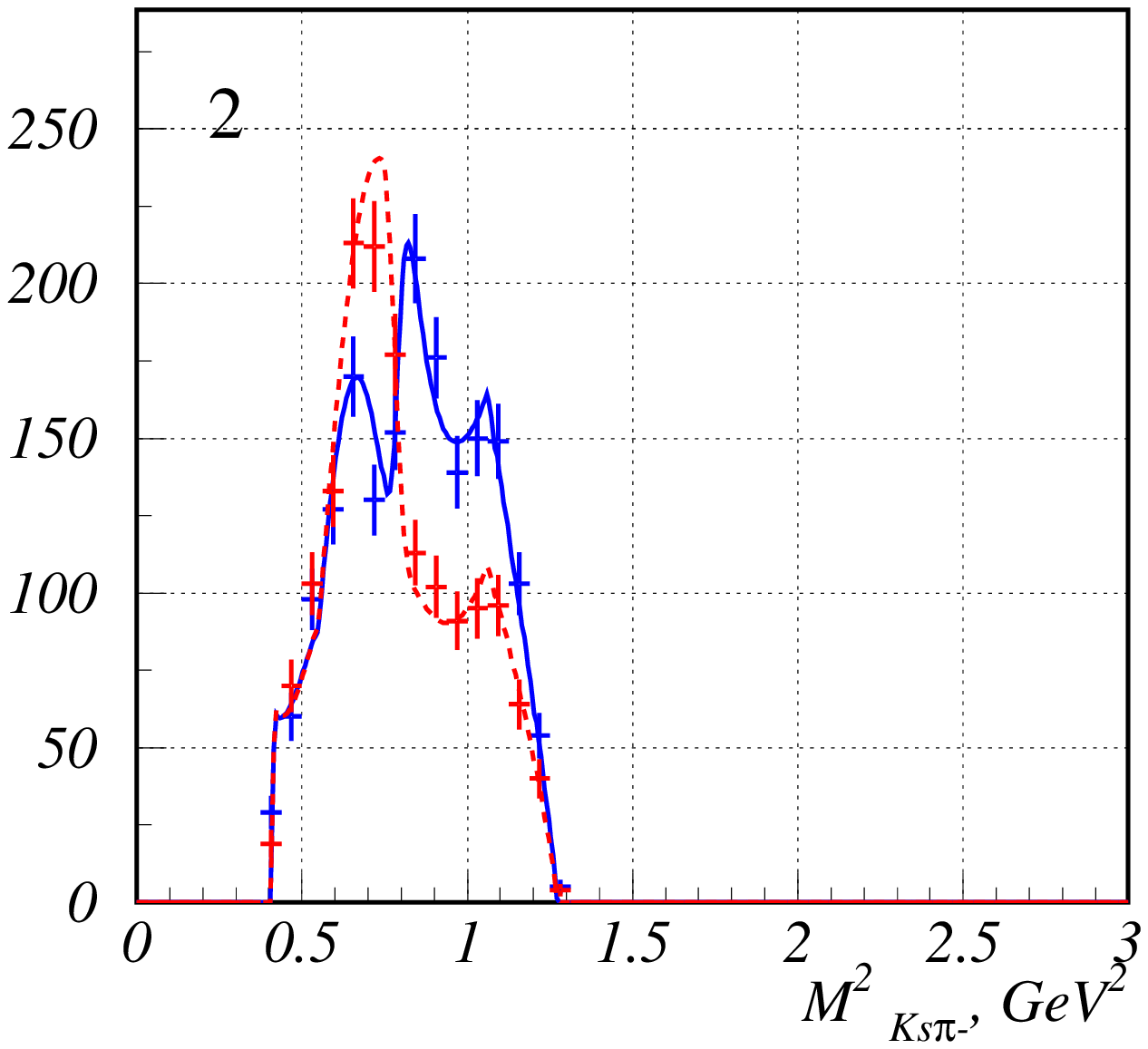,width=0.45\textwidth}
  \vspace{-0.05\textwidth}
  \epsfig{figure=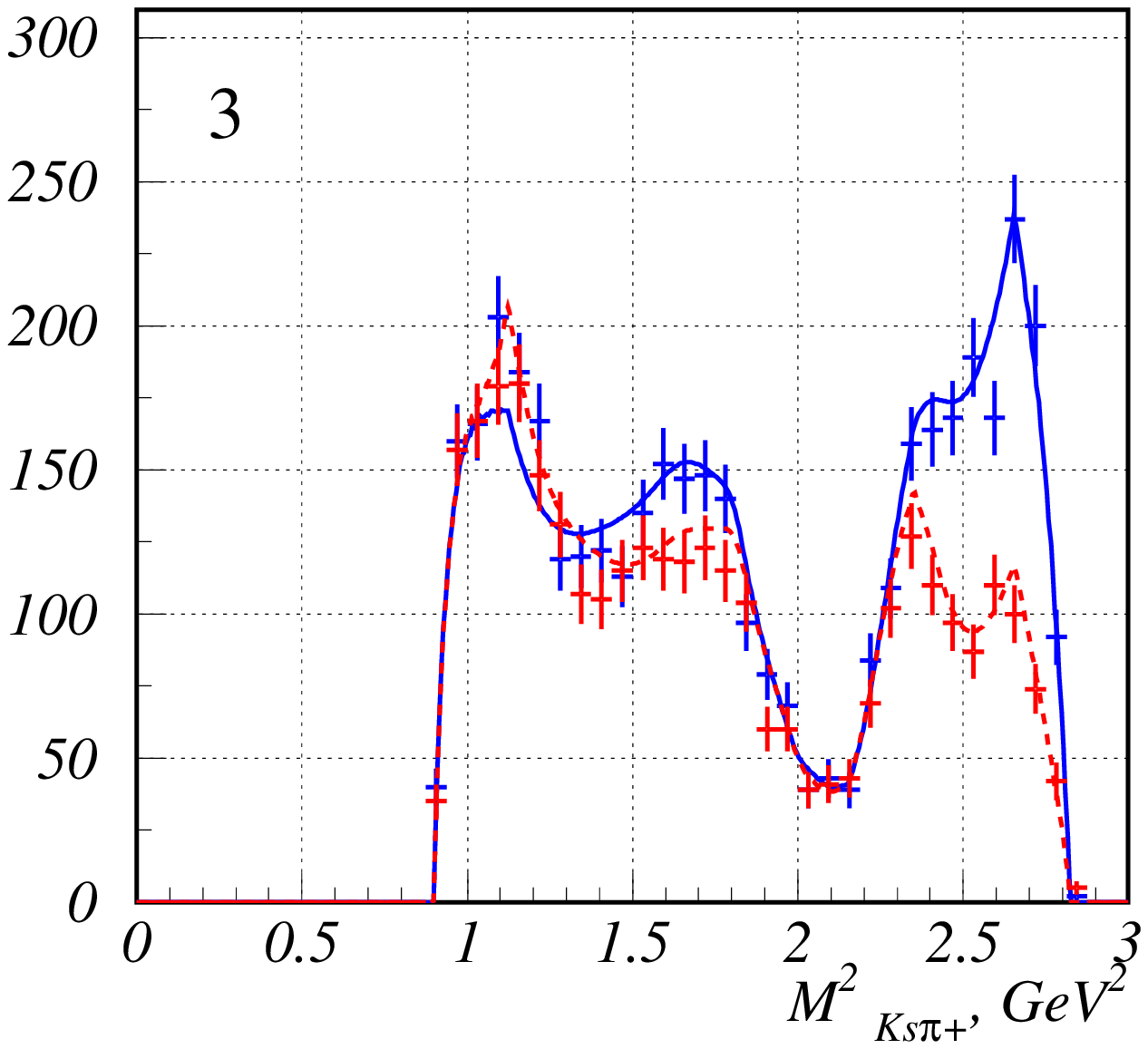,width=0.45\textwidth}
  \hspace{-0.05\textwidth}
  \epsfig{figure=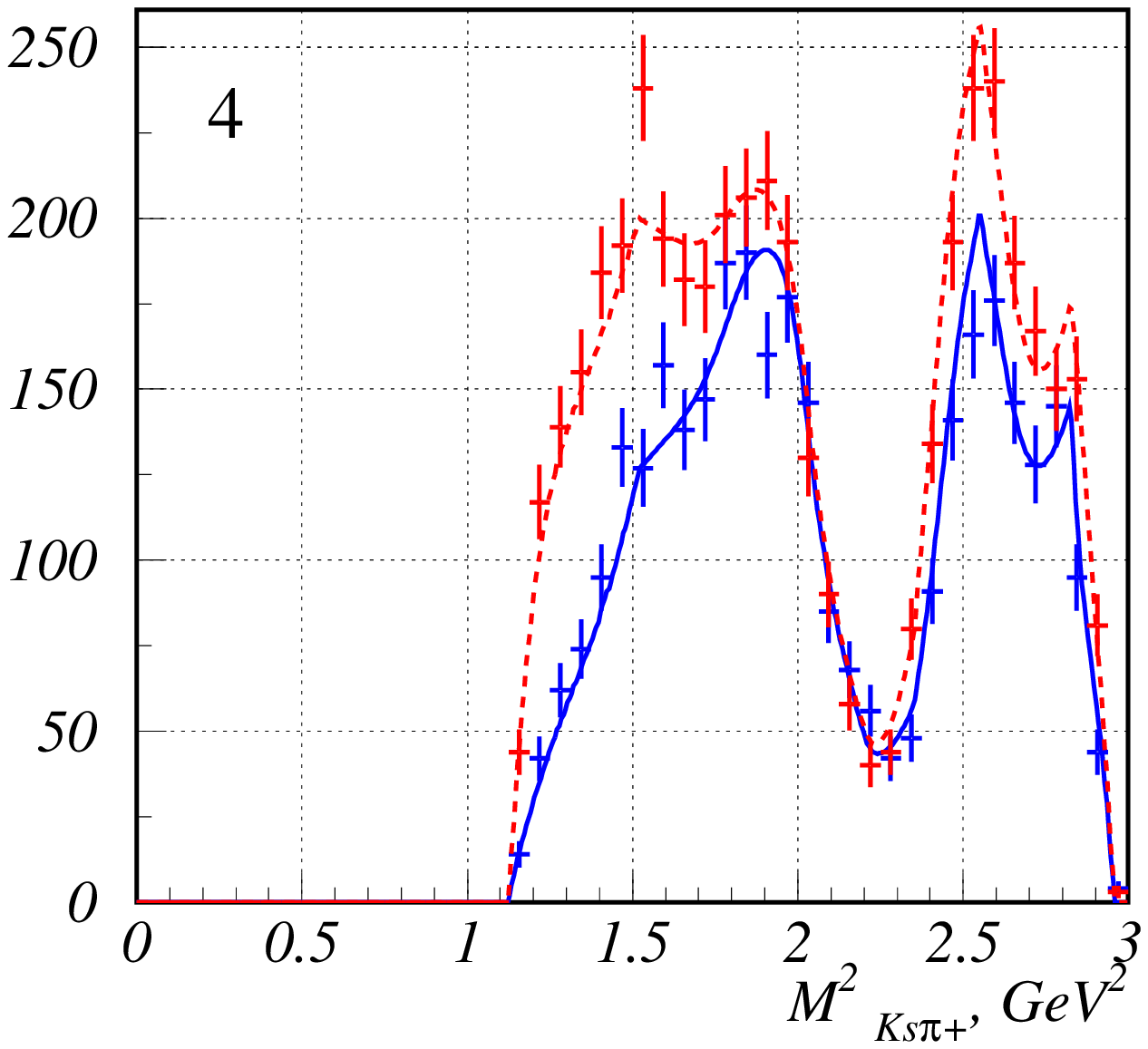,width=0.45\textwidth}
  \caption{Simulated Dalitz plot of $D^0$ decay and its slices 
           showing the effect of CP asymmetry in $B\to D^0 K$ decay with 
           the parameters $a=0.125$, $\phi_3=70^{\circ}$ and $\delta=0$. 
           The two curves represent the Dalitz plot density for
           $D^0$ decay from $B^+\to \bar{D^0} K^+$ (solid line) and $B^-\to D^0 K^-$
           (dashed line) processes. 
           }
  \label{slices}
  \end{center}
\end{figure}

\begin{figure}
  \vspace{-0.05\textwidth}
  \begin{center}
  \epsfig{figure=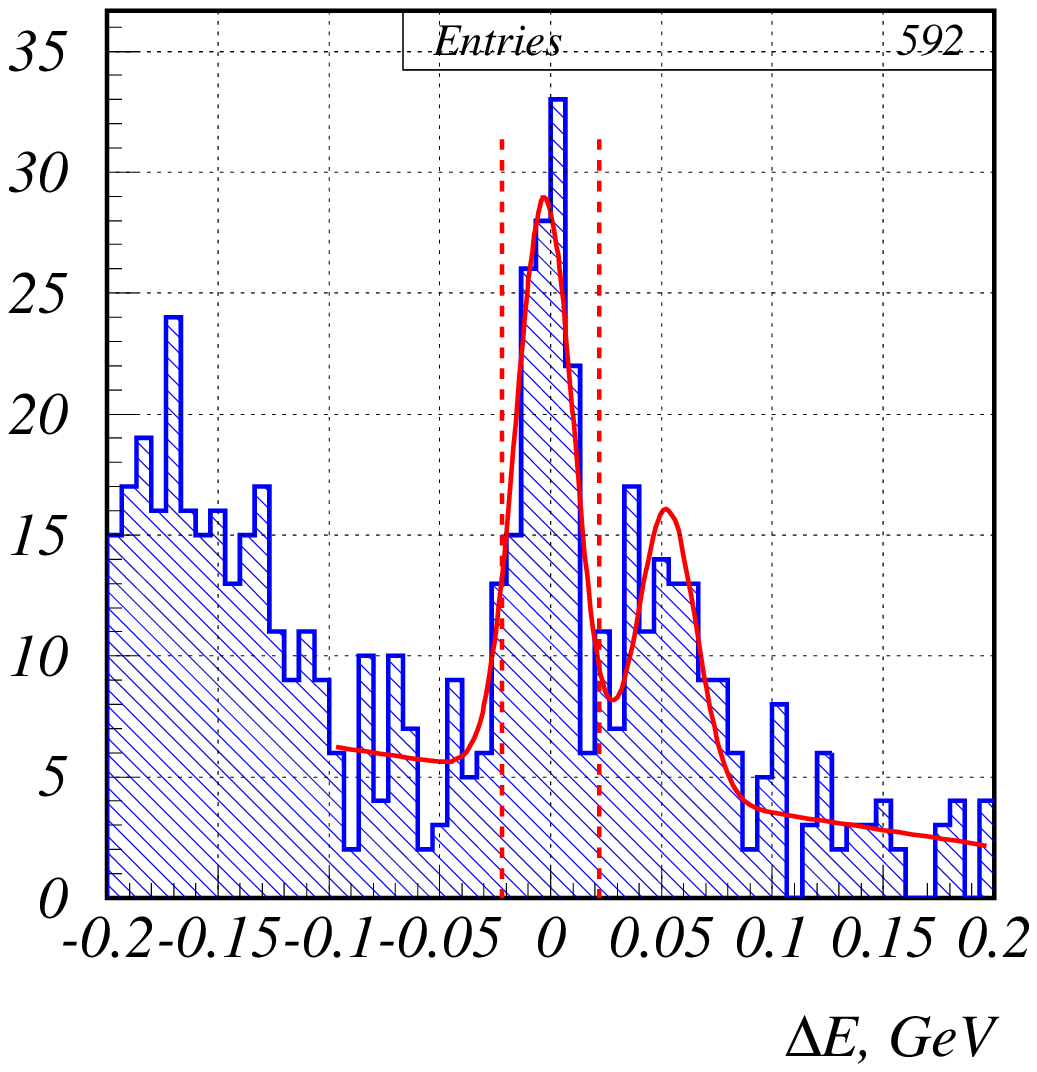,width=0.35\textwidth}
  \hspace{-0.05\textwidth}
  \epsfig{figure=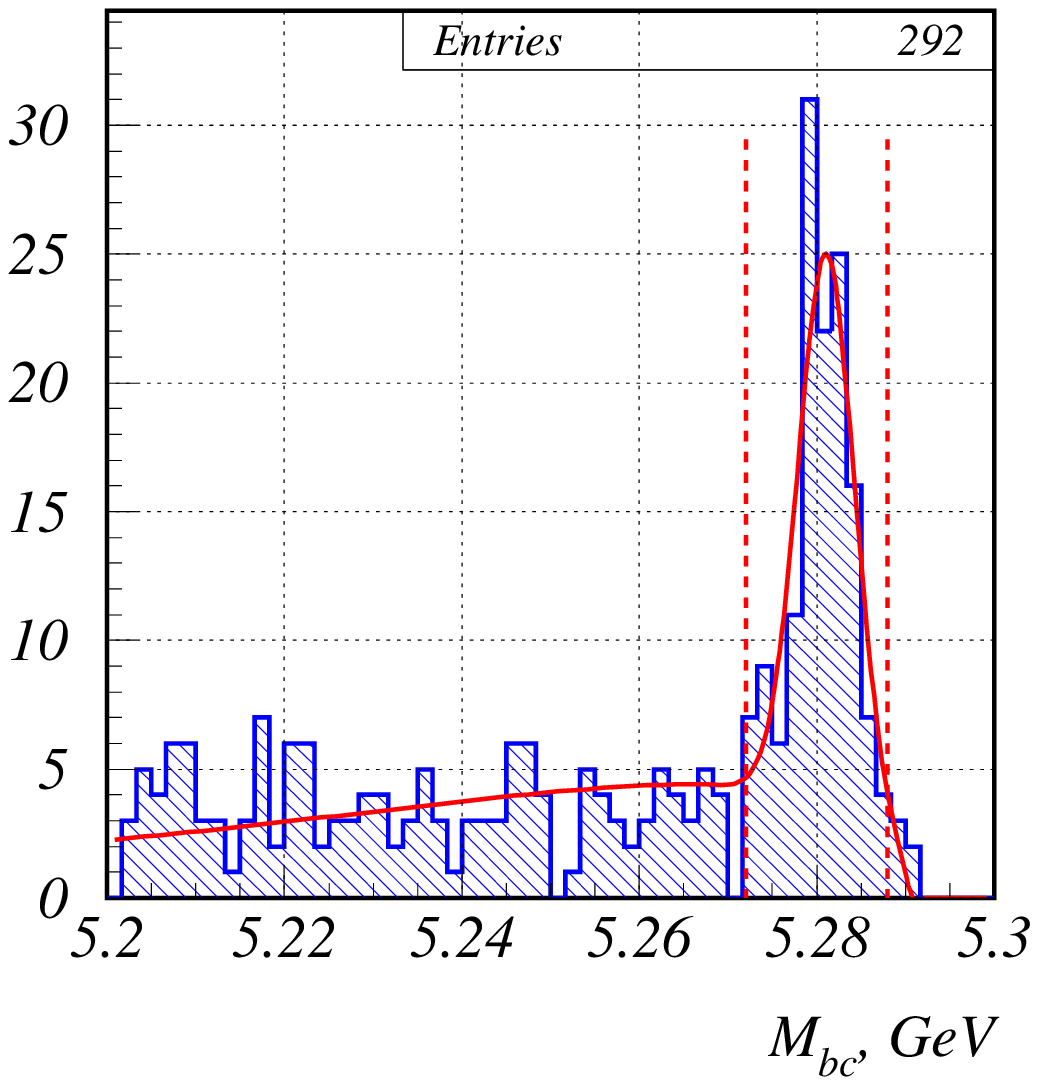,width=0.35\textwidth}
  \hspace{-0.05\textwidth}
  \epsfig{figure=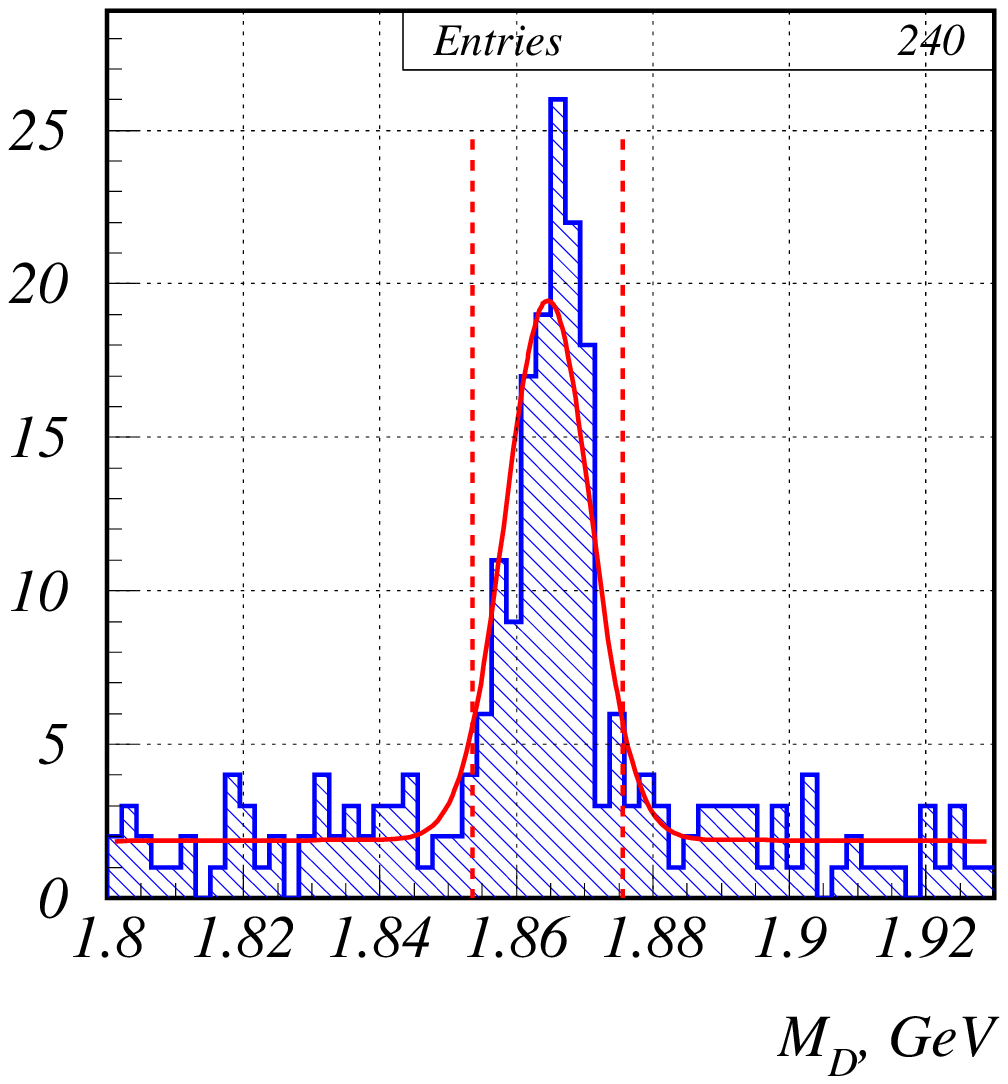,width=0.35\textwidth}
  \caption{$\Delta E$, $M_{bc}$ and $M_D$ distributions for $B\to D^0 K$ decay. 
  Histogram represents the data, the solid line is the 
  fit result, dashed lines show the cut limits for the corresponding
  distributions.}
  \label{b2dk}
  \vspace{\baselineskip}
  \end{center}
\end{figure}

\begin{figure}
  \vspace{-0.05\textwidth}
  \begin{center}
  \epsfig{figure=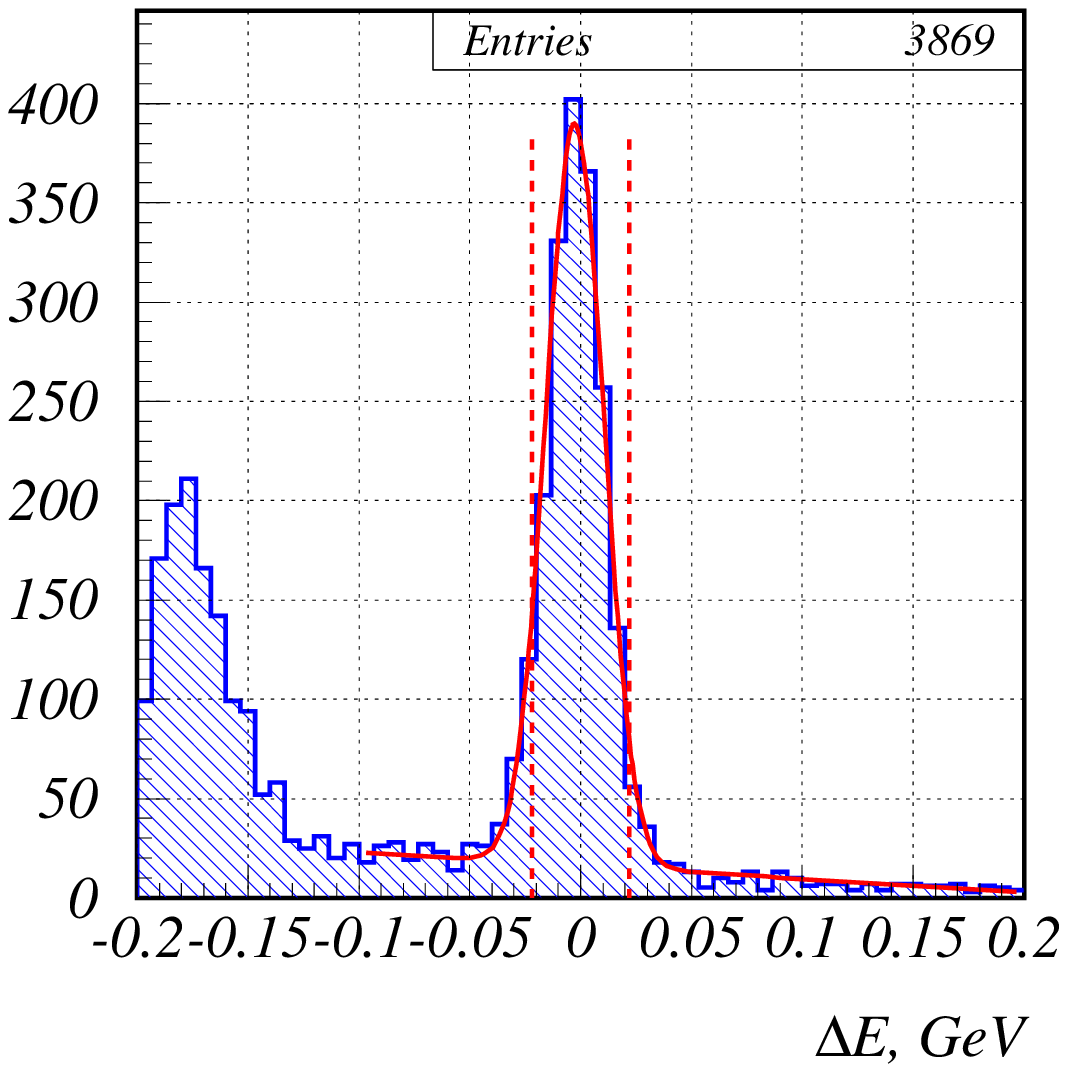,width=0.35\textwidth}
  \hspace{-0.05\textwidth}
  \epsfig{figure=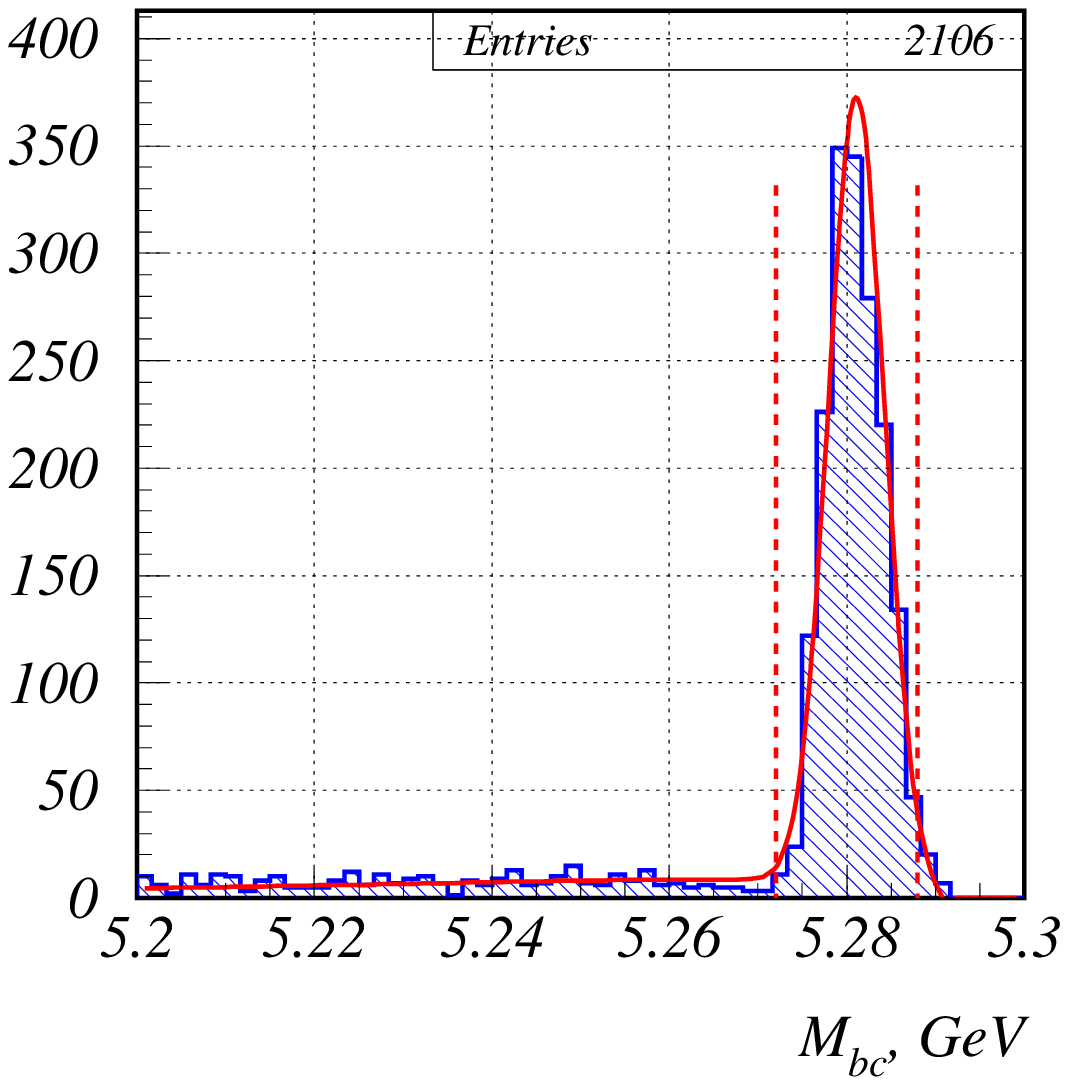,width=0.35\textwidth}
  \hspace{-0.05\textwidth}
  \epsfig{figure=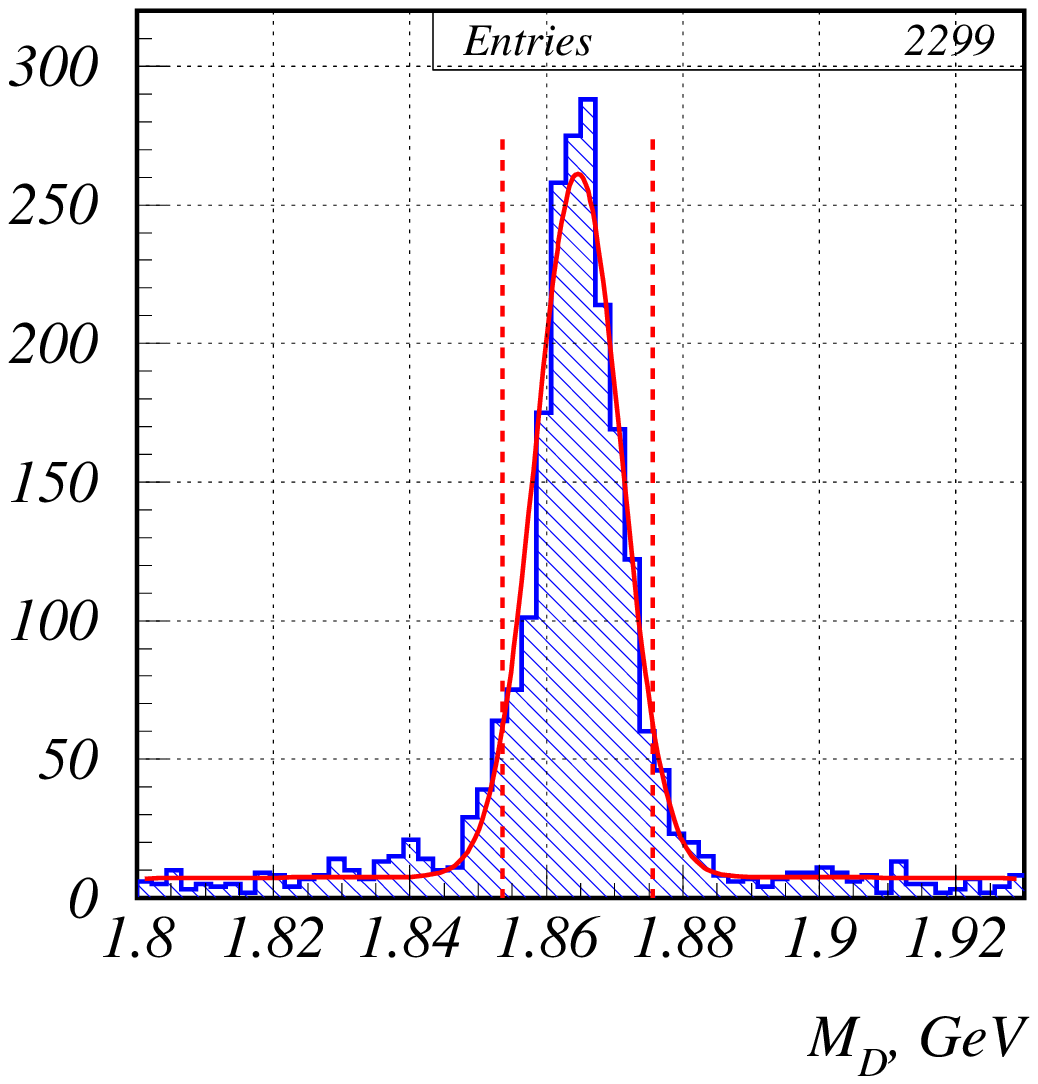,width=0.35\textwidth}
  \caption{$\Delta E$, $M_{bc}$ and $M_D$ distributions for $B\to D^0\pi$ decay.}
  \label{b2dpi}
  \vspace{\baselineskip}
  \end{center}
\end{figure}

\begin{figure}
  \vspace{-0.05\textwidth}
  \begin{center}
  \epsfig{figure=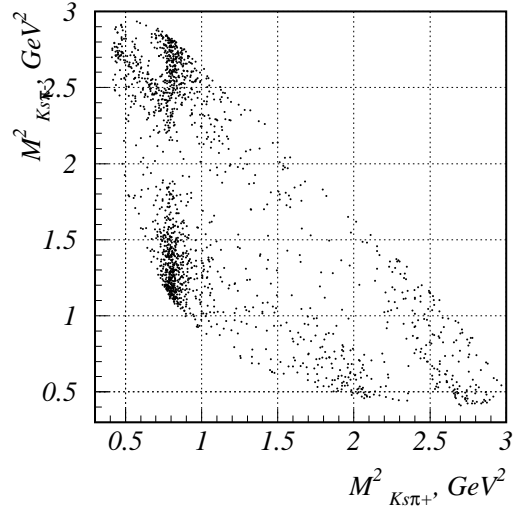,width=0.48\textwidth}
  \caption{Dalitz distribution
   in variables $m^2_{K_s\pi^+}$, $m^2_{K_s\pi^-}$ of 
   $D^0\to K_s\pi^+\pi^-$ decay from $B\to D^0 \pi$ process.}
  \label{b2dpi_plot}
  \end{center}
\end{figure}

\begin{figure}
  \centering
  \epsfig{figure=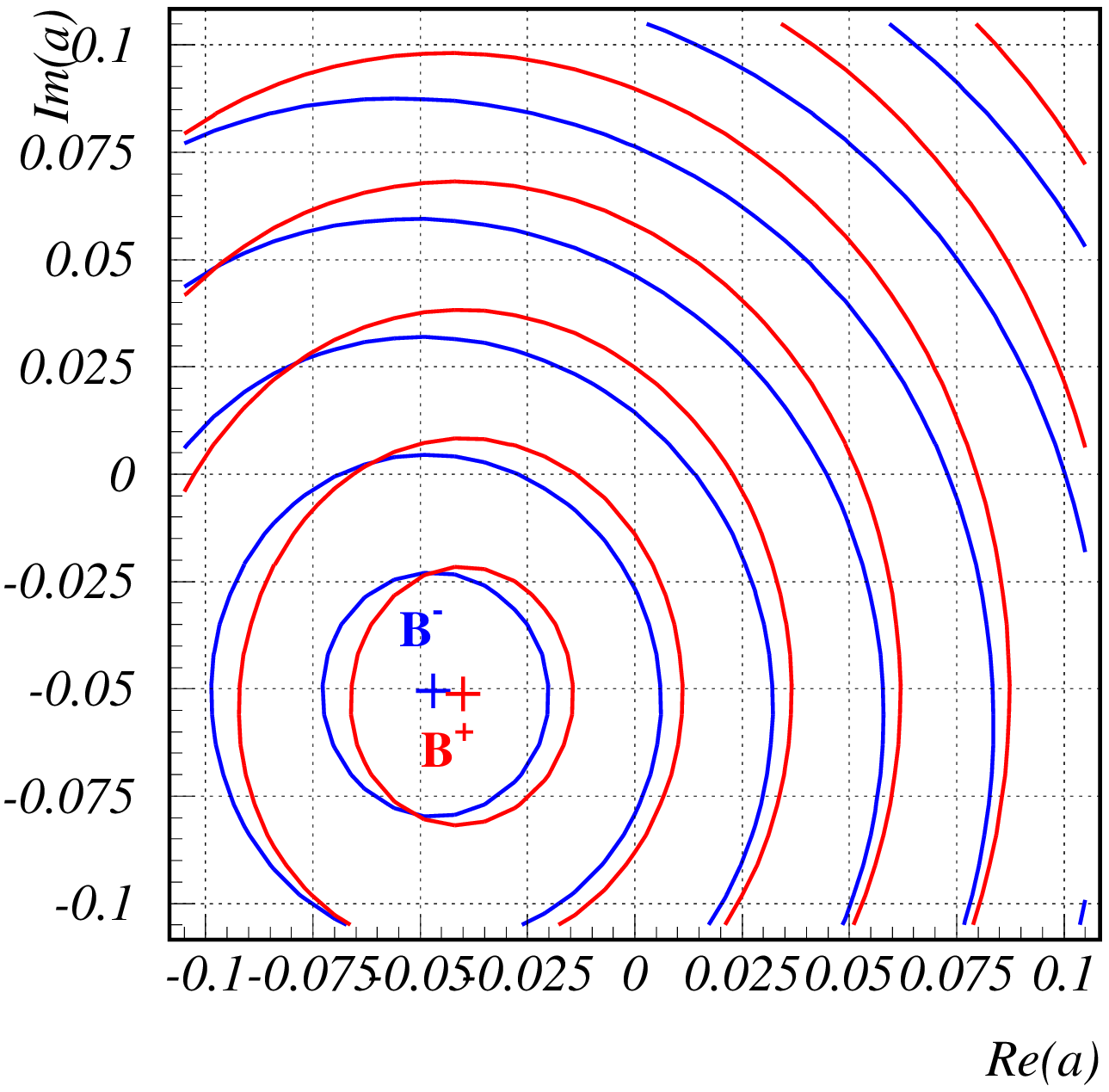,width=0.45\textwidth}
  \epsfig{figure=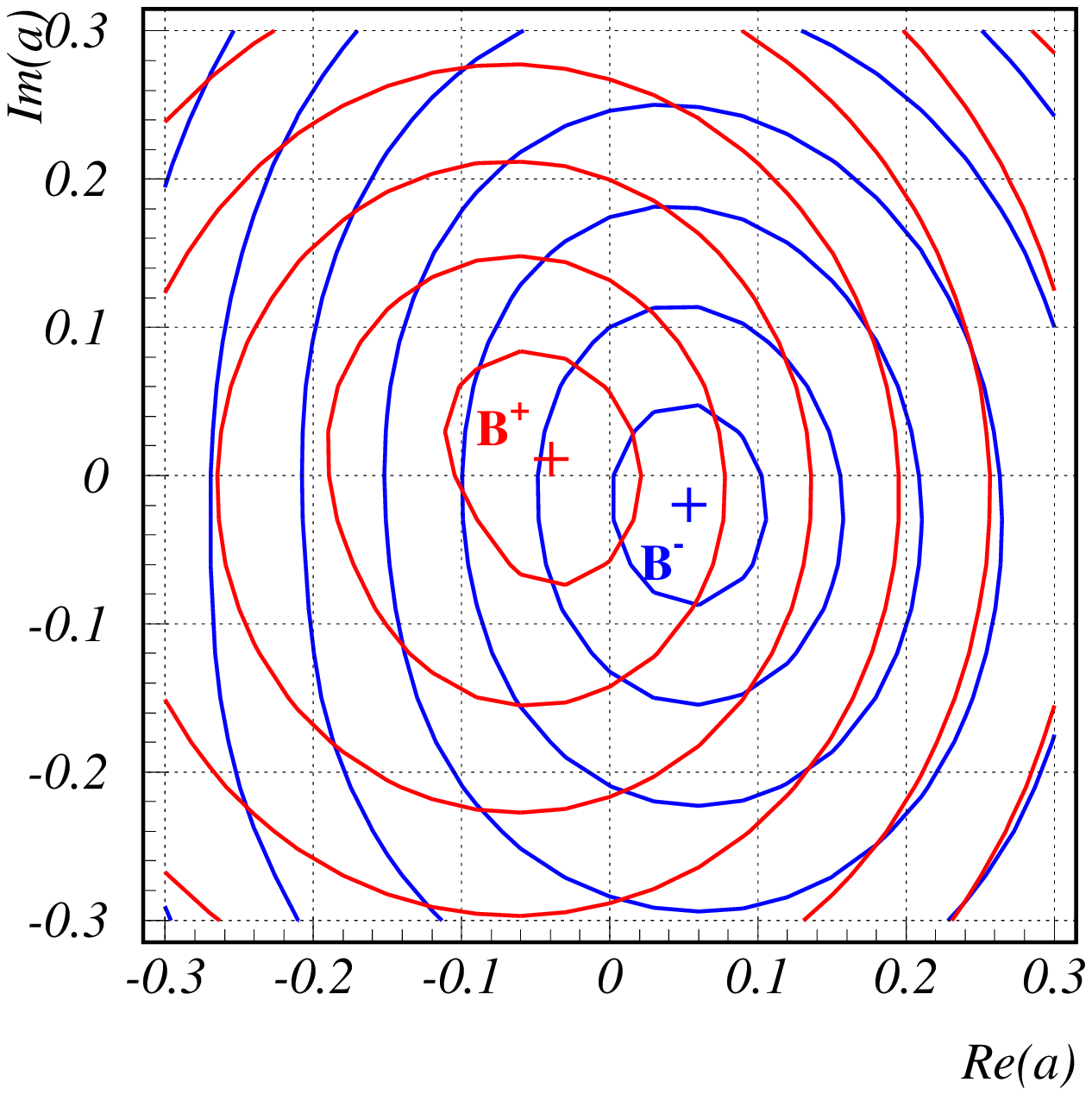,width=0.45\textwidth}

  \caption{Constraint plots of complex relative amplitude 
           $ae^{i\theta}$ for $B\to D^0\pi$ (left) and $B\to D^{*0}\pi$ (right) 
	   decays. Contours indicate integer multiples of the standard deviation.}
  \label{constr_plot}
\end{figure}

\begin{figure}
  \vspace{-0.05\textwidth}
  \begin{center} 

  \epsfig{figure=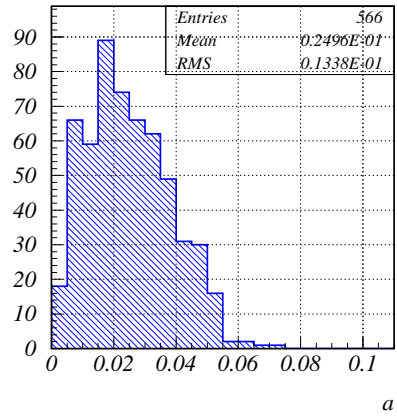,width=0.4\textwidth}

  \caption{Distribution of the fitted parameter $a$ in 566 trials
           of 1700 events each, generated with $a = 0$.}
  \label{a_mc}
  \end{center} 
\end{figure}

\begin{figure}
  \vspace{-0.05\textwidth}
  \begin{center}
  \epsfig{figure=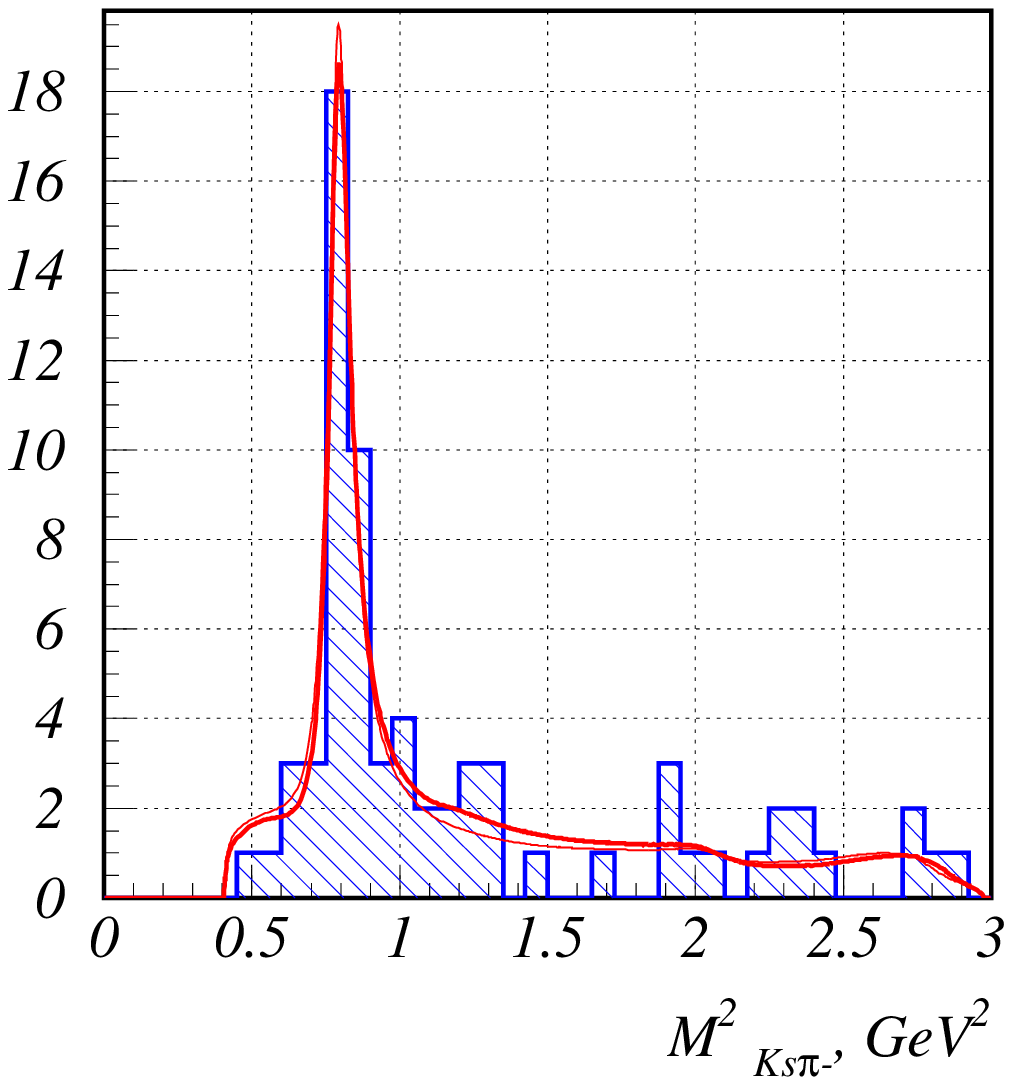,width=0.45\textwidth}
  \hspace{-0.05\textwidth}
  \epsfig{figure=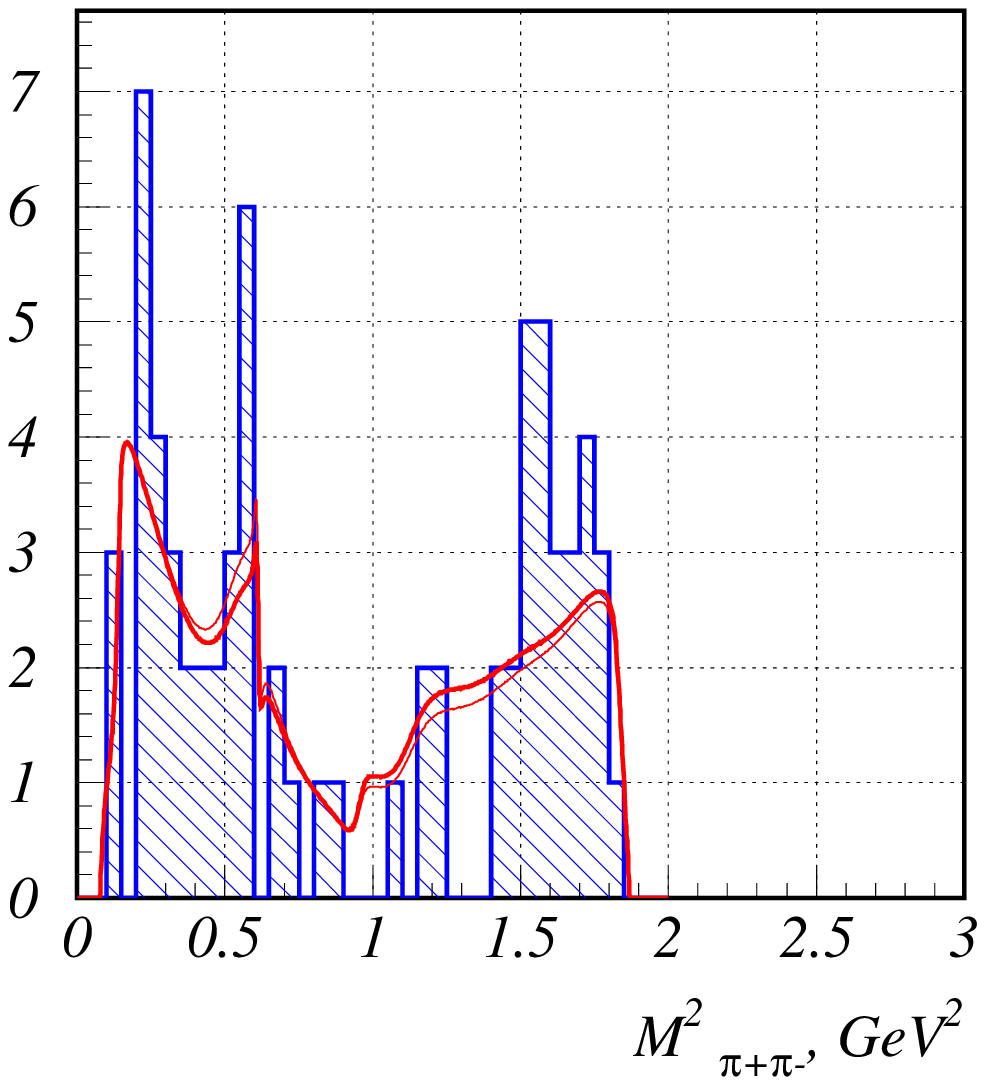,width=0.45\textwidth}
  \epsfig{figure=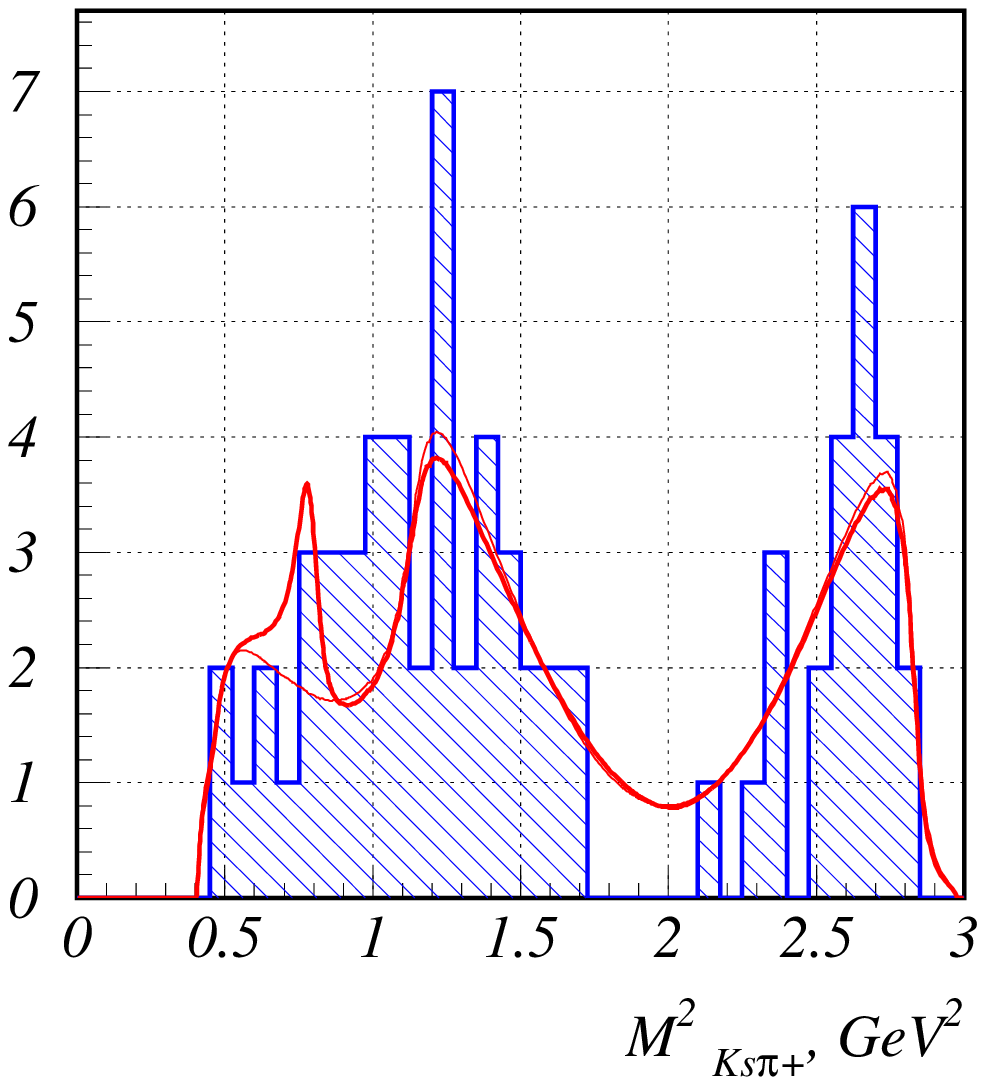,width=0.45\textwidth}
  \hspace{-0.05\textwidth}
  \epsfig{figure=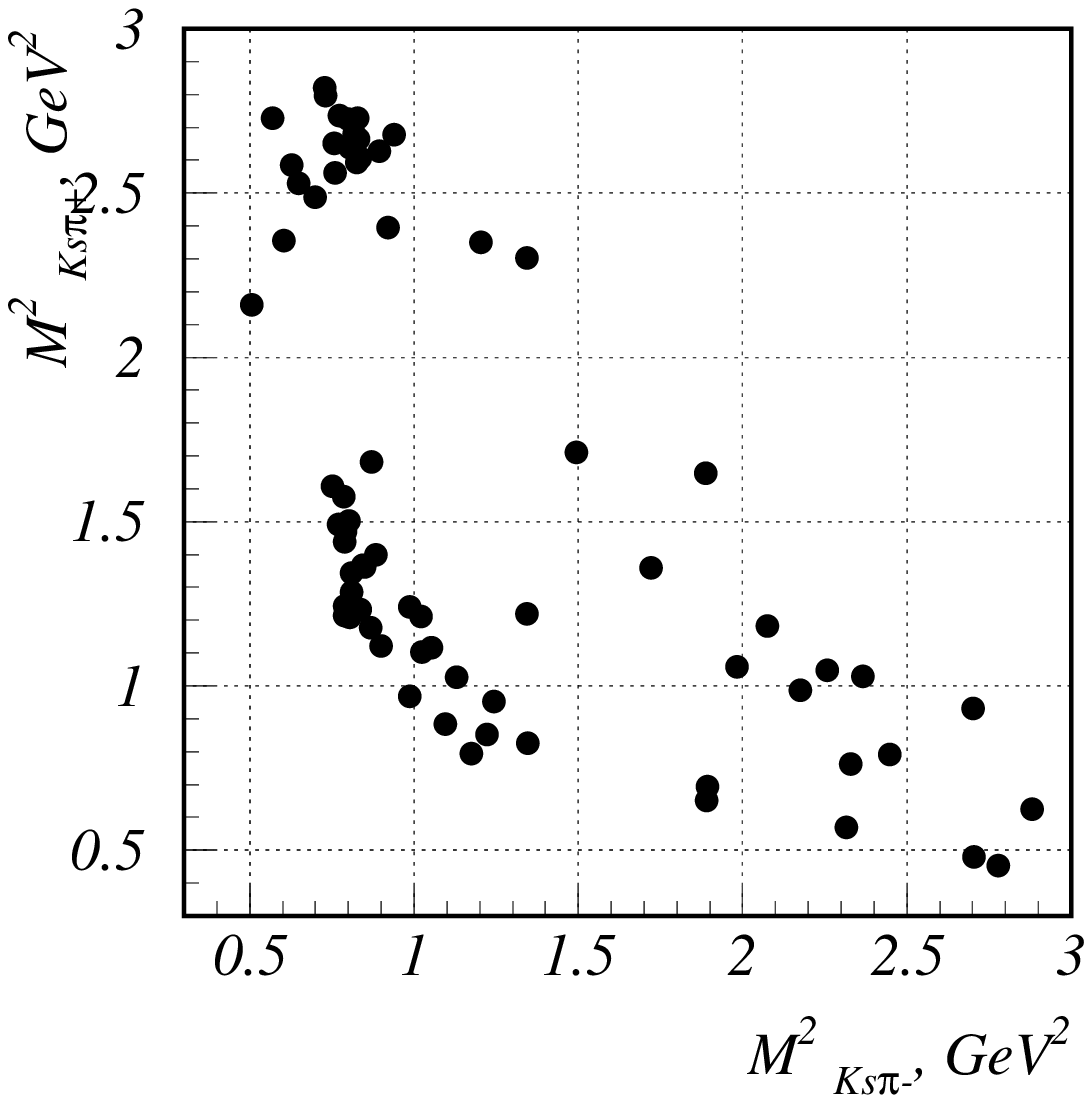,width=0.45\textwidth}
  \caption{$M^2_{K\pi^-}$, $M^2_{K\pi^+}$, $M^2_{\pi^+\pi^-}$ distributions
   and Dalitz plot of $D^0\to K_s\pi^+\pi^-$ decay from $B^-\to D^0 K^-$
   process. Thick line represents the fit result, thin line is $a=0$ case. }
  \label{b2dk_plot_m}
  \end{center}
\end{figure}

\begin{figure}
  \vspace{-0.05\textwidth}
  \begin{center}
  \epsfig{figure=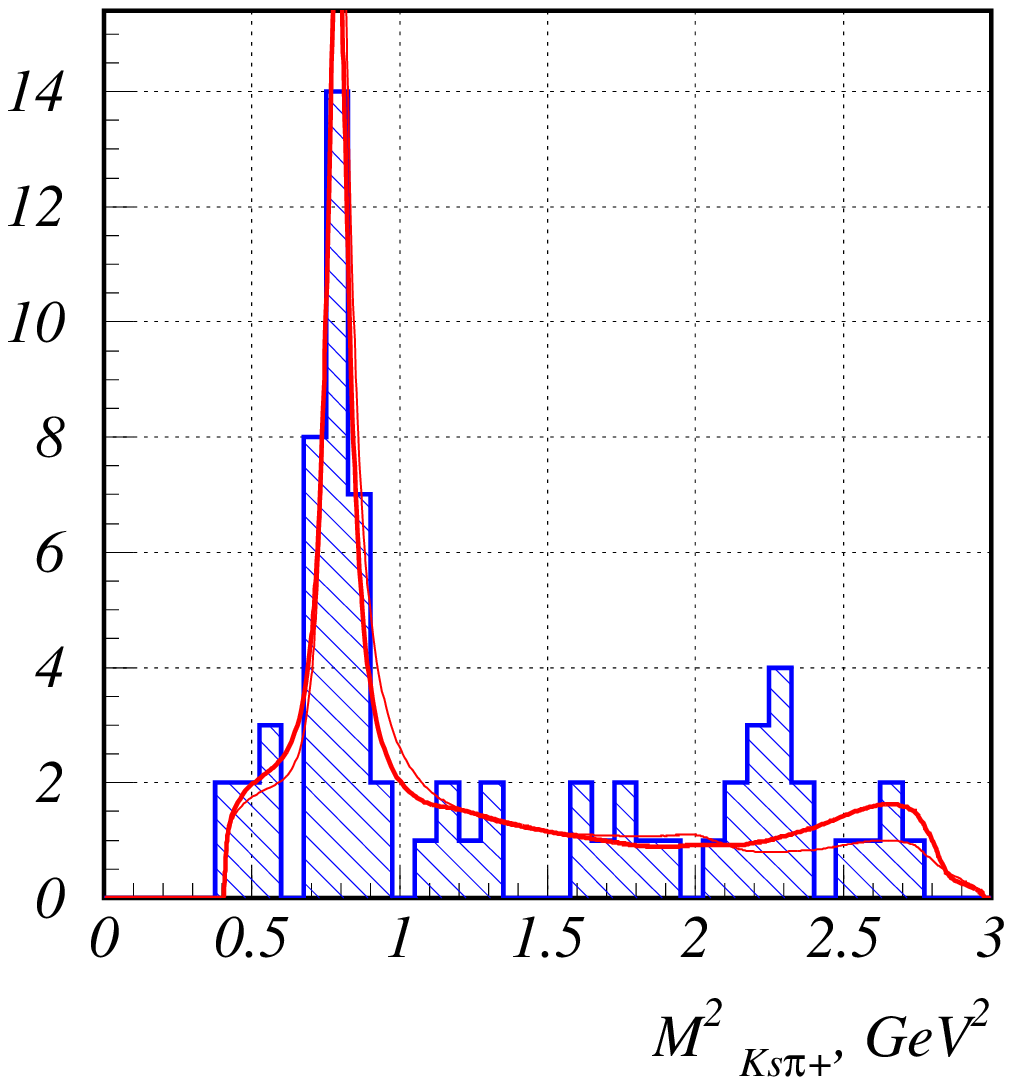,width=0.45\textwidth}
  \hspace{-0.05\textwidth}
  \epsfig{figure=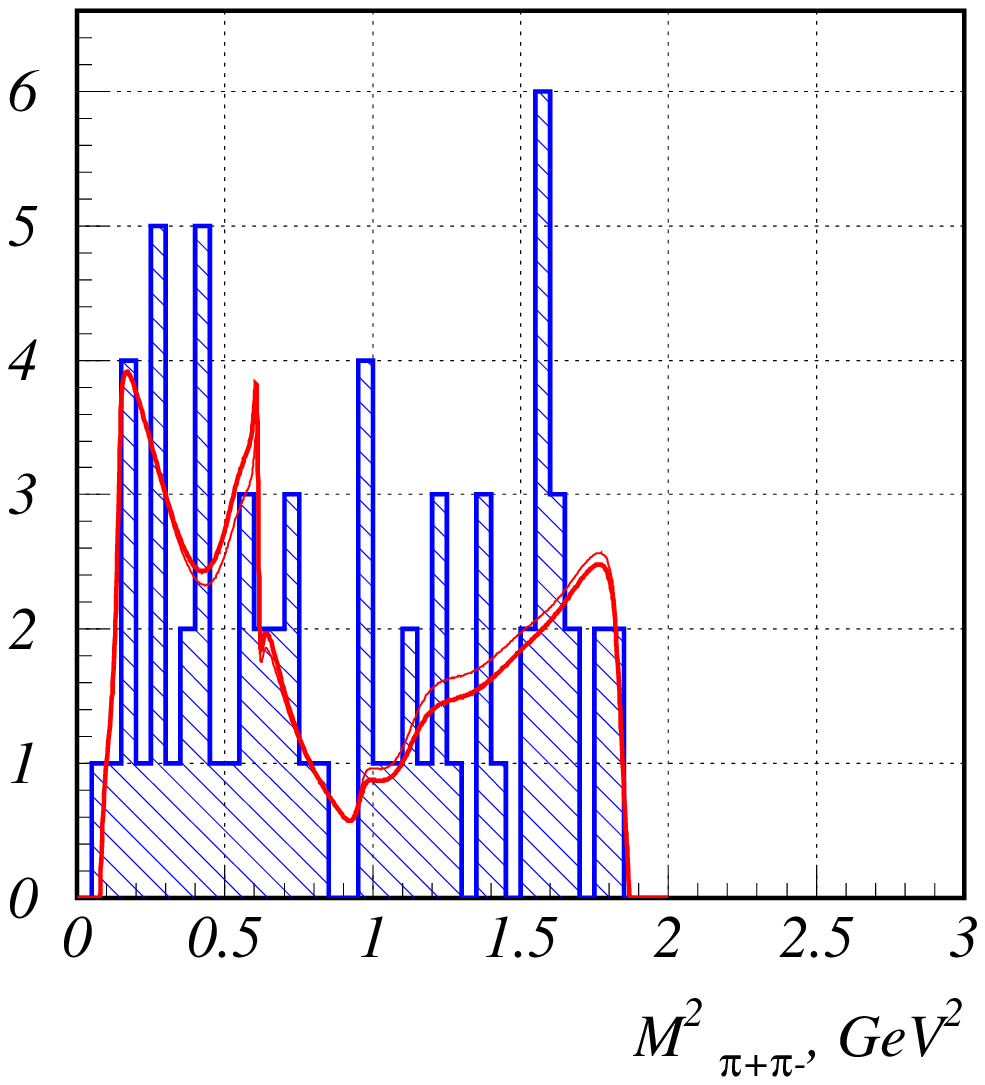,width=0.45\textwidth}
  \epsfig{figure=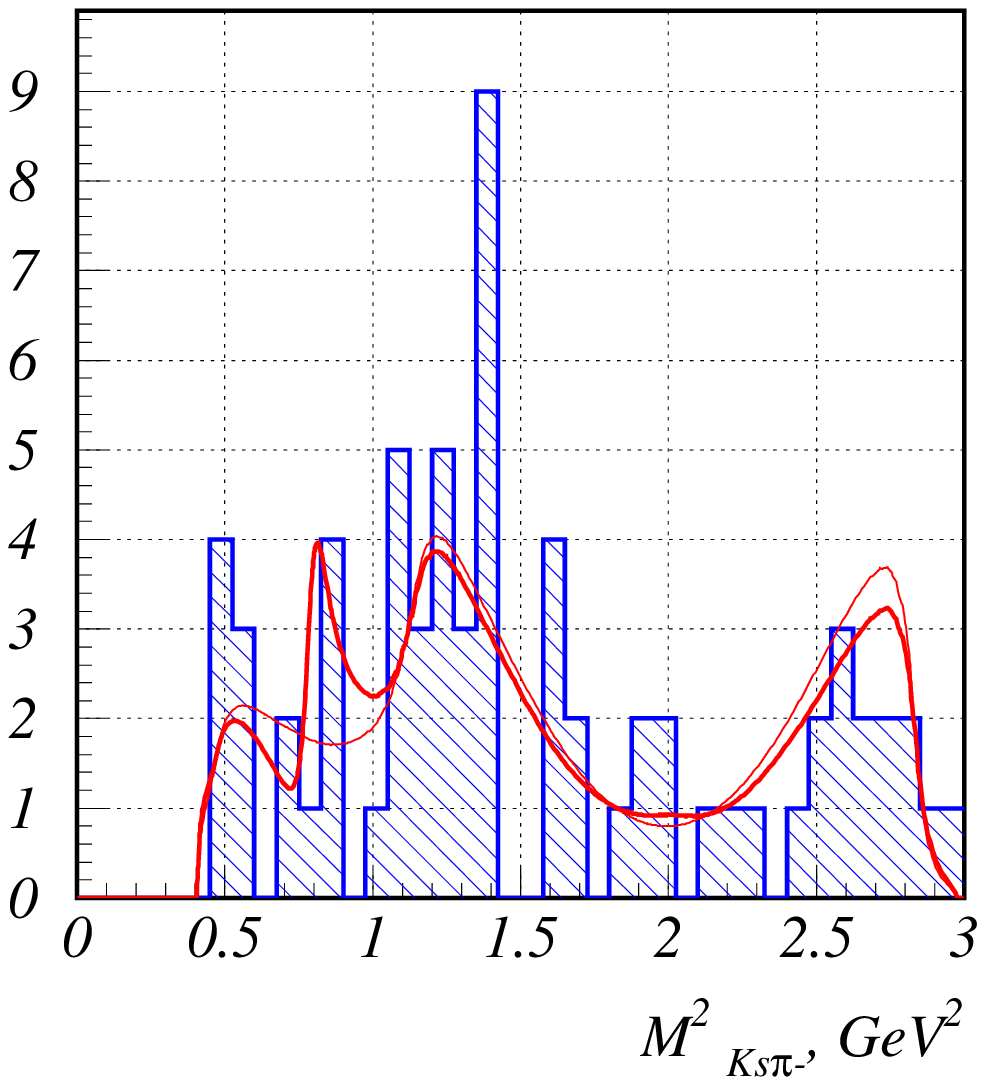,width=0.45\textwidth}
  \hspace{-0.05\textwidth}
  \epsfig{figure=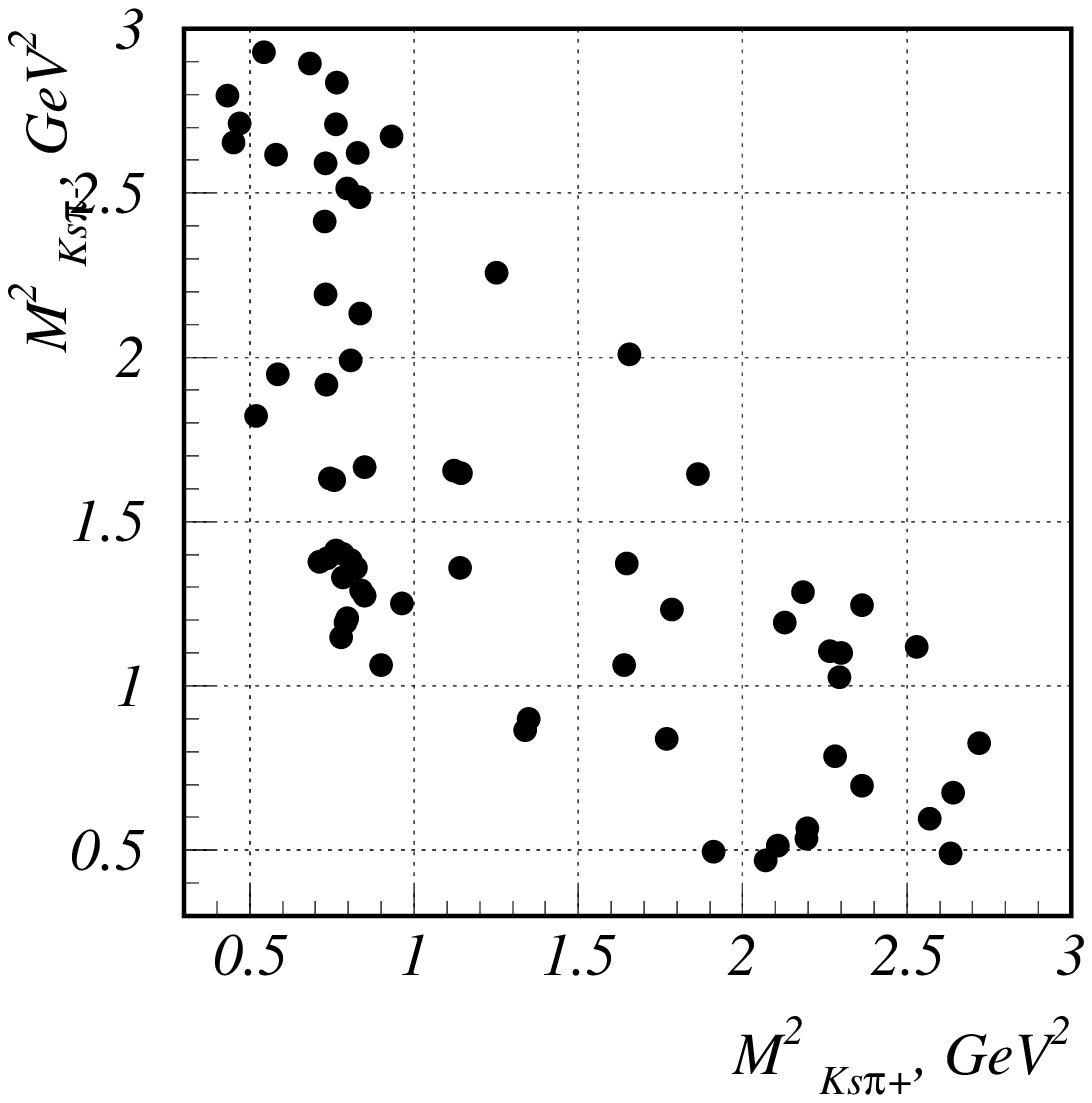,width=0.45\textwidth}
  \caption{$M^2_{K\pi^+}$, $M^2_{K\pi^-}$, $M^2_{\pi^+\pi^-}$ distributions
   and Dalitz plot of $D^0\to K_s\pi^+\pi^-$ decay from $B^+\to \bar{D^0}K^+$
   process. Thick line represents the fit result, thin line is $a=0$ case. }
  \label{b2dk_plot_p}
  \end{center}
\end{figure}

\begin{figure}
  \vspace{-0.05\textwidth}
  \begin{center} 

  \epsfig{figure=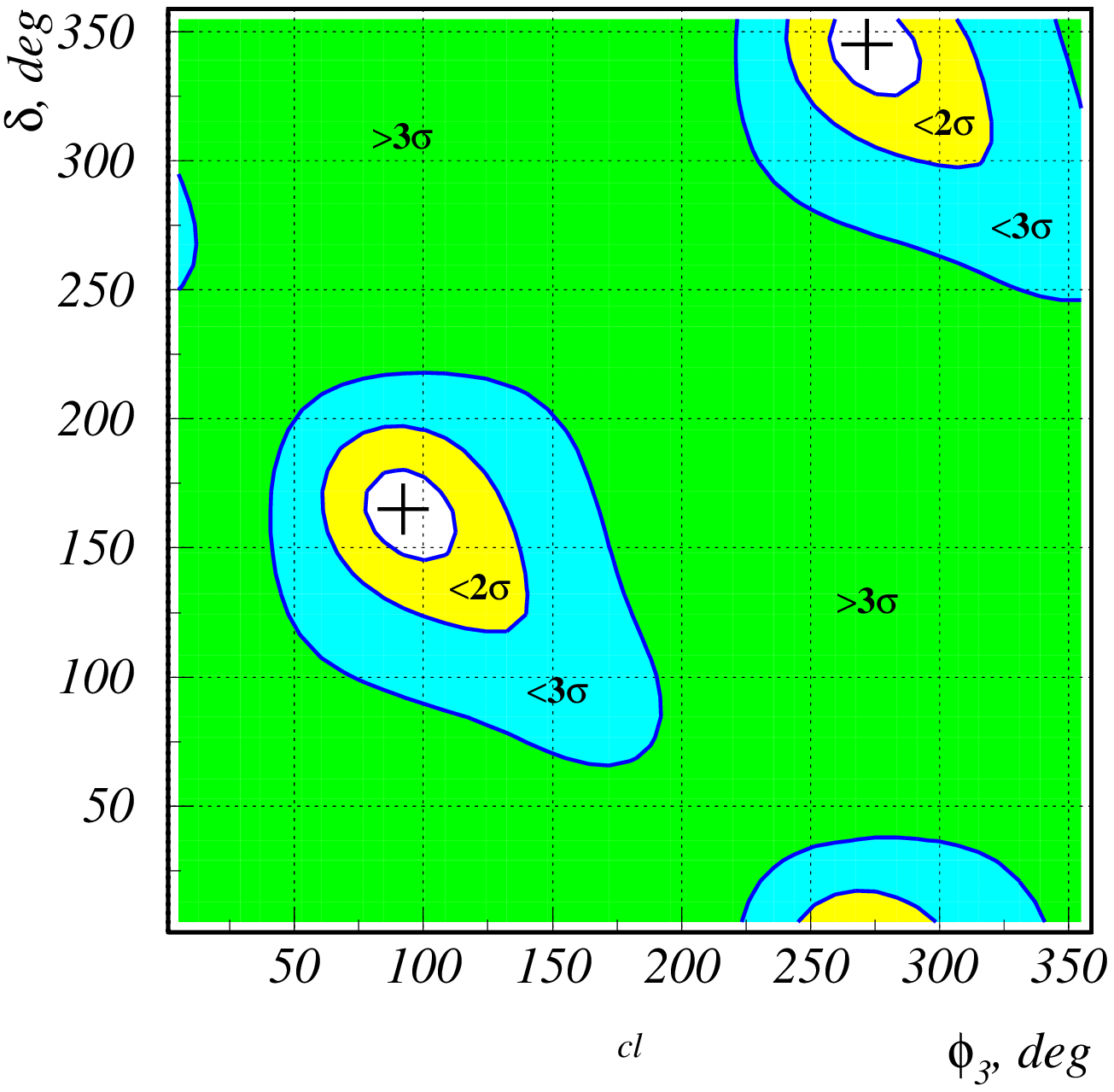,width=0.47\textwidth}
  \hfill
  \epsfig{figure=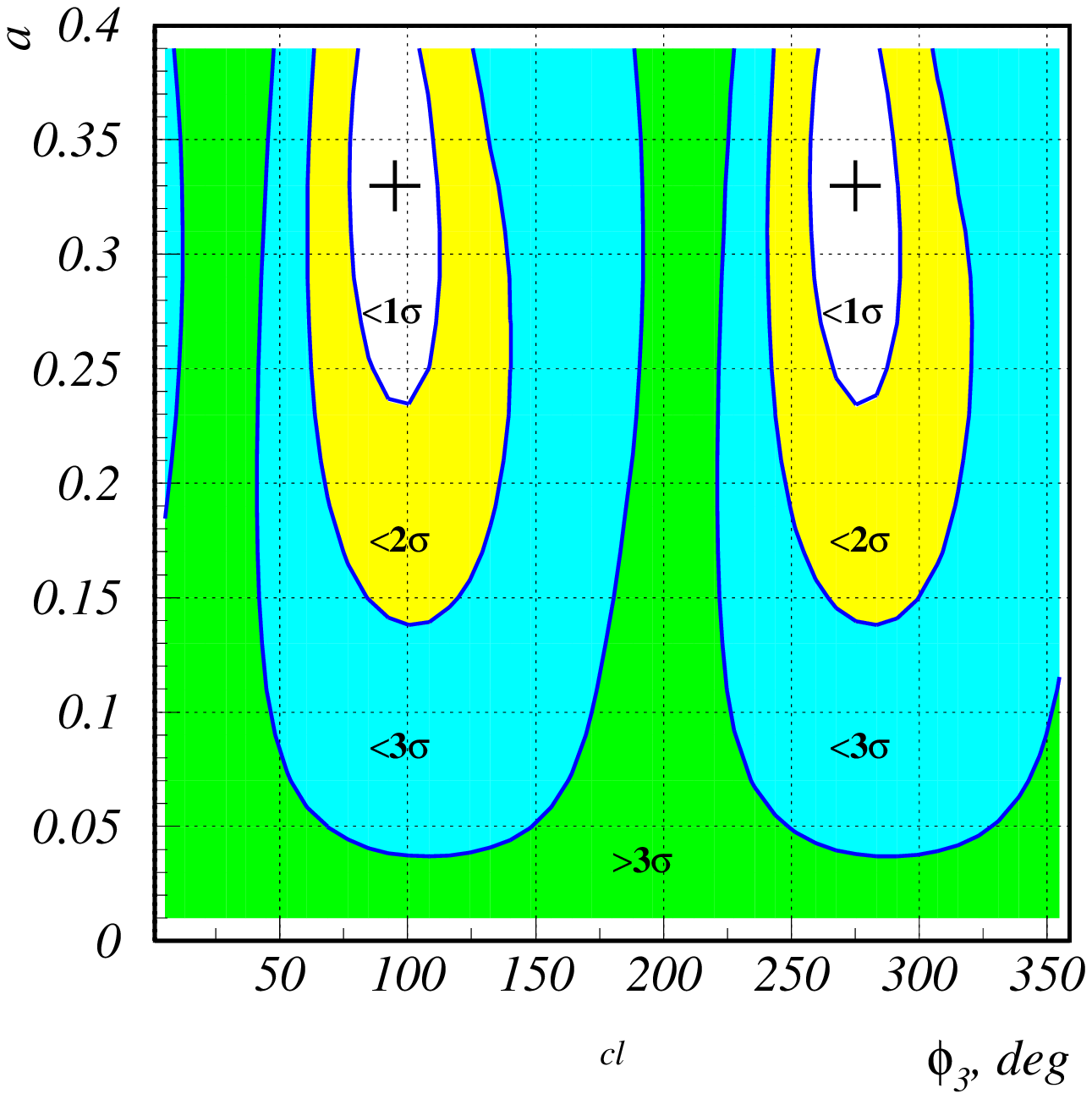,width=0.47\textwidth}

  \caption{Constraints on $\phi_3$ and $\delta$ (left) and 
           $a$ and $\phi_3$ (right) for all parameters free in the fit.}
  \label{b2dk_free}

  \epsfig{figure=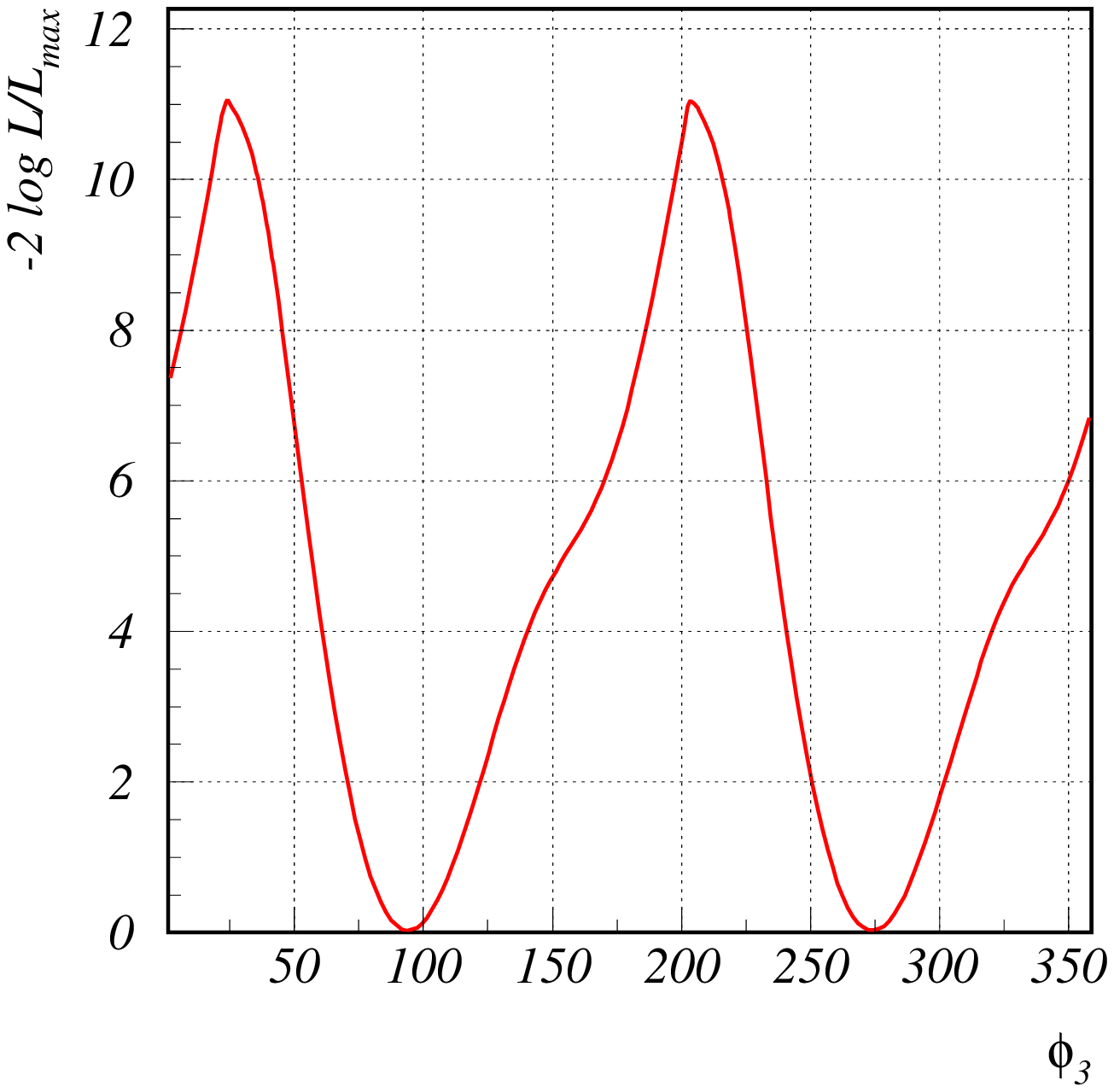,width=0.47\textwidth}
  \hfill
  \epsfig{figure=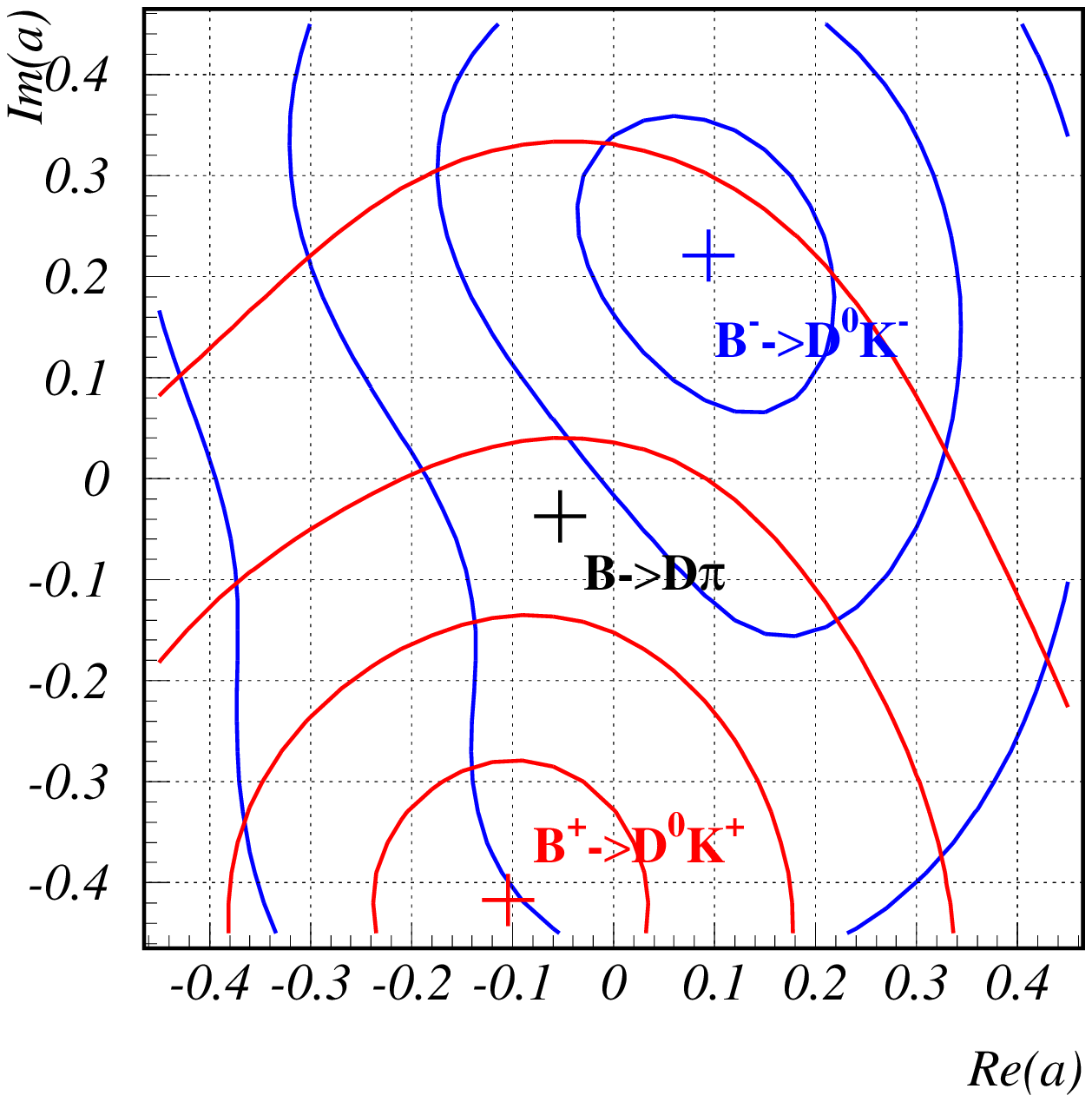,width=0.47\textwidth}

  \caption{$\phi_3$ statistical significance (left) and results of the 
           complex relative amplitude $a\exp(i\theta)$ fits for $B^+\to \bar{D^0}\pi^+$
           and $B^-\to D^0\pi^-$ samples (right). 
           Contours indicate integer multiples of the standard deviation.}
  \label{b2dk_asym}

  \end{center}
\end{figure}

\begin{figure}
  \epsfig{figure=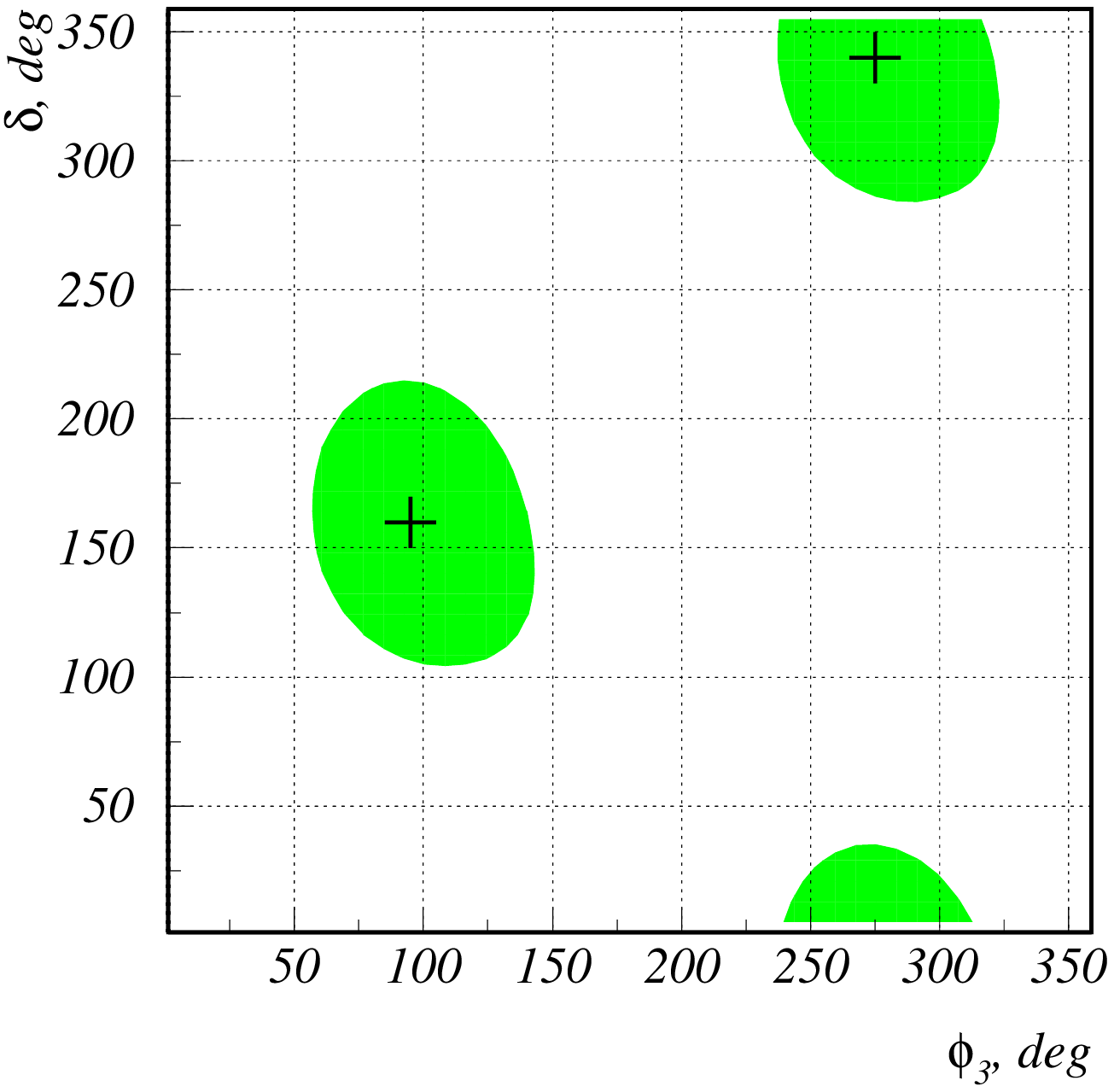,width=0.47\textwidth}
  \hfill
  \epsfig{figure=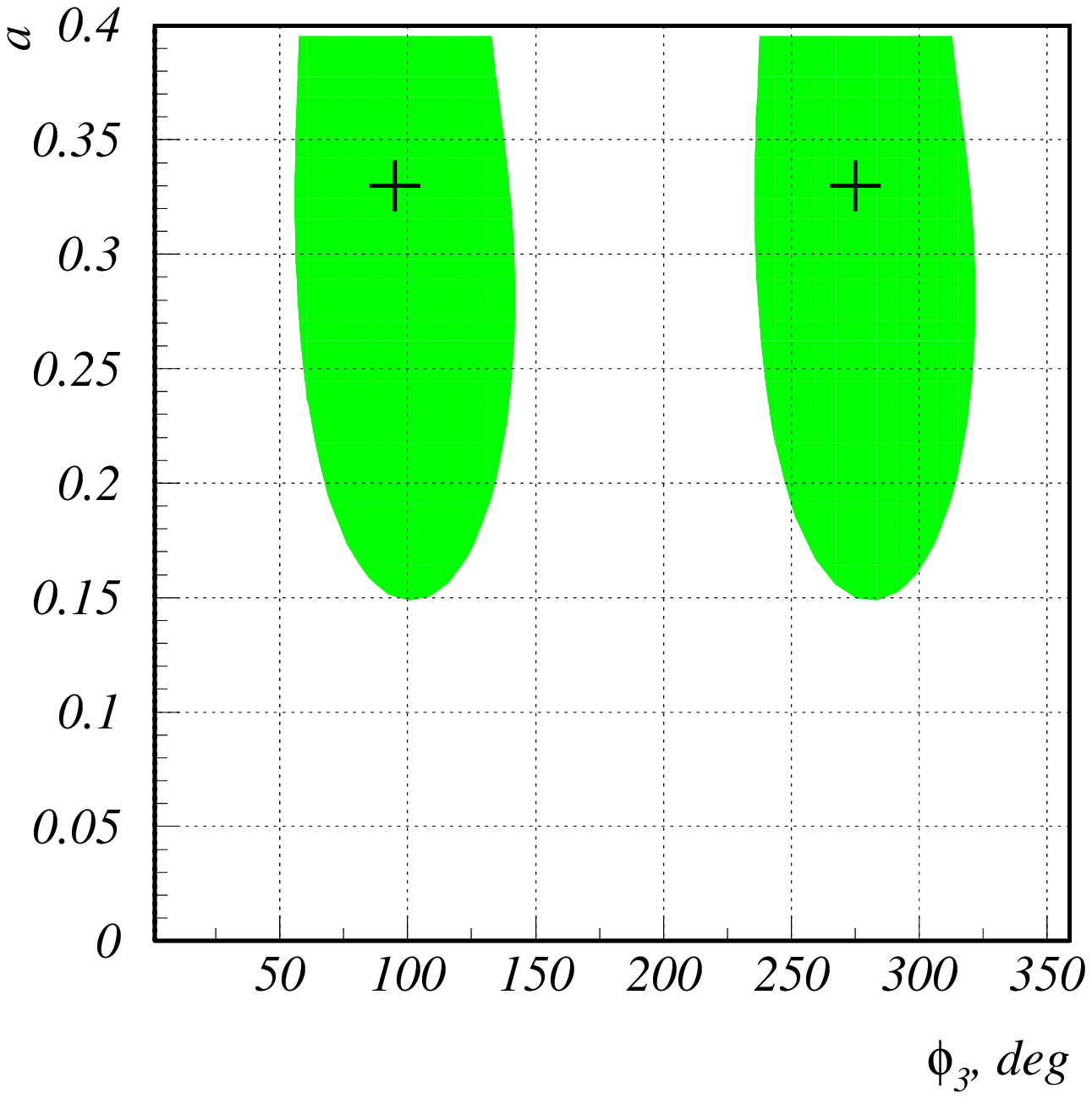,width=0.47\textwidth}

  \caption{90\% CL regions for $\phi_3$ and $\delta$ (left) and 
           $a$ and $\phi_3$ (right) with systematic uncertainty included.}
  \label{b2dk_90cl}
\end{figure}

\end{document}

%% file: author.tex
\affiliation{Aomori University, Aomori}
\affiliation{Budker Institute of Nuclear Physics, Novosibirsk}
\affiliation{Chiba University, Chiba}
\affiliation{Chuo University, Tokyo}
\affiliation{University of Cincinnati, Cincinnati, Ohio 45221}
\affiliation{University of Frankfurt, Frankfurt}
\affiliation{Gyeongsang National University, Chinju}
\affiliation{University of Hawaii, Honolulu, Hawaii 96822}
\affiliation{High Energy Accelerator Research Organization (KEK), Tsukuba}
\affiliation{Hiroshima Institute of Technology, Hiroshima}
\affiliation{Institute of High Energy Physics, Chinese Academy of Sciences, Beijing}
\affiliation{Institute of High Energy Physics, Vienna}
\affiliation{Institute for Theoretical and Experimental Physics, Moscow}
\affiliation{J. Stefan Institute, Ljubljana}
\affiliation{Kanagawa University, Yokohama}
\affiliation{Korea University, Seoul}
\affiliation{Kyoto University, Kyoto}
\affiliation{Kyungpook National University, Taegu}
\affiliation{Institut de Physique des Hautes \'Energies, Universit\'e de Lausanne, Lausanne}
\affiliation{University of Ljubljana, Ljubljana}
\affiliation{University of Maribor, Maribor}
\affiliation{University of Melbourne, Victoria}
\affiliation{Nagoya University, Nagoya}
\affiliation{Nara Women's University, Nara}
\affiliation{National Kaohsiung Normal University, Kaohsiung}
\affiliation{National Lien-Ho Institute of Technology, Miao Li}
\affiliation{Department of Physics, National Taiwan University, Taipei}
\affiliation{H. Niewodniczanski Institute of Nuclear Physics, Krakow}
\affiliation{Nihon Dental College, Niigata}
\affiliation{Niigata University, Niigata}
\affiliation{Osaka City University, Osaka}
\affiliation{Osaka University, Osaka}
\affiliation{Panjab University, Chandigarh}
\affiliation{Peking University, Beijing}
\affiliation{Princeton University, Princeton, New Jersey 08545}
\affiliation{RIKEN BNL Research Center, Upton, New York 11973}
\affiliation{Saga University, Saga}
\affiliation{University of Science and Technology of China, Hefei}
\affiliation{Seoul National University, Seoul}
\affiliation{Sungkyunkwan University, Suwon}
\affiliation{University of Sydney, Sydney NSW}
\affiliation{Tata Institute of Fundamental Research, Bombay}
\affiliation{Toho University, Funabashi}
\affiliation{Tohoku Gakuin University, Tagajo}
\affiliation{Tohoku University, Sendai}
\affiliation{Department of Physics, University of Tokyo, Tokyo}
\affiliation{Tokyo Institute of Technology, Tokyo}
\affiliation{Tokyo Metropolitan University, Tokyo}
\affiliation{Tokyo University of Agriculture and Technology, Tokyo}
\affiliation{Toyama National College of Maritime Technology, Toyama}
\affiliation{University of Tsukuba, Tsukuba}
\affiliation{Utkal University, Bhubaneswer}
\affiliation{Virginia Polytechnic Institute and State University, Blacksburg, Virginia 24061}
\affiliation{Yokkaichi University, Yokkaichi}
\affiliation{Yonsei University, Seoul}
  \author{K.~Abe}\affiliation{High Energy Accelerator Research Organization (KEK), Tsukuba} 
  \author{K.~Abe}\affiliation{Tohoku Gakuin University, Tagajo} 
  \author{N.~Abe}\affiliation{Tokyo Institute of Technology, Tokyo} 
  \author{R.~Abe}\affiliation{Niigata University, Niigata} 
  \author{T.~Abe}\affiliation{High Energy Accelerator Research Organization (KEK), Tsukuba} 
  \author{I.~Adachi}\affiliation{High Energy Accelerator Research Organization (KEK), Tsukuba} 
  \author{Byoung~Sup~Ahn}\affiliation{Korea University, Seoul} 
  \author{H.~Aihara}\affiliation{Department of Physics, University of Tokyo, Tokyo} 
  \author{M.~Akatsu}\affiliation{Nagoya University, Nagoya} 
  \author{M.~Asai}\affiliation{Hiroshima Institute of Technology, Hiroshima} 
% ...
\collaboration{The Belle Collaboration}

%% file: bc0343.bbl
\begin{thebibliography}{100}

\bibitem{ckm}
M. Kobayashi and T. Maskawa, Prog. Theor. Phys. {\bf 49}, 652 (1973); 
N. Cabibbo, Phys. Rev. Lett. 10, 531 (1963);

\bibitem{bdk_gamma}
M. Gronau and D. Wyler, Phys. Lett. {\bf B265}, 172 (1991);
I. Dunietz, Phys. Lett. {\bf B270}, 75 (1991);
D. Atwood, G. Eilam, M. Gronau and A. Soni, Phys. Lett. {\bf B341}, 372 (1995);
D. Atwood, I. Dunietz and A. Soni, Phys. Rev. Lett. {\bf 78}, 3257 (1997).

\bibitem{dalitz_bdk}
A. Giri, Yu. Grossman, A. Soffer, J. Zupan, hep-ph/0303187.

\bibitem{bdk_cleo}
CLEO Collaboration, M. Athanas {\it et al.}, Phys. Rev. Lett. {\bf 80}, 5493 (1998).

\bibitem{bdk_color_supp}
Belle Collaboration, P. Krokovny {\it et al.}, Phys. Rev. Lett. {\bf 90}, 141802
(2003).

\bibitem{bdk_color_all}
Belle Collaboration, K. Abe {\it et al.}, Phys. Rev. Lett. {\bf 87}, 111801 (2001).

\bibitem{gronau}
M. Gronau, Phys. Lett. B {\bf 557}, 198 (2003)

\bibitem{dkpp_cleo}
CLEO Collaboration, H. Muramatsu {\it et al.}, Phys. Rev. Lett.
{\bf 89}, 251802 (2002), Erratum-ibid: {\bf 90}, 059901 (2003).

\bibitem{cleo_model}
CLEO Collaboration, S. Kopp {\it et al.}, Phys. Rev. D {\bf 63}, 092001 (2001).

\bibitem{belle}
Belle Collaboration, A.~Abashian {\it et al.}, Nucl. Instr. and Meth. A {\bf 479}, 117 (2002).

\bibitem{fisher}
CLEO Collaboration, D. M. Asner {\it et al.}, Phys. Rev. D {\bf 53}, 1039 (1996)

\bibitem{argus}
ARGUS collaboration, H. Albrecht {\it et al.}, Phys. Lett. B {\bf 241}, 278 (1990). 

\end{thebibliography}
